\documentclass [twocolumn,amsmath,amssymb,showpacs]{revtex4}
\usepackage{subfigure}
\usepackage{wrapfig}
\usepackage{color}
\usepackage[dvipdfmx]{graphicx}
\usepackage{dcolumn}
\usepackage{bm}
\usepackage{amsmath}
\usepackage{amsfonts}
\usepackage{amssymb}
\usepackage{url}
\allowdisplaybreaks[0]
\makeatletter

\begin{document}
\title{Granular rotor as a probe for a non-equilibrium bath}

\author{Tomohiko G. Sano$^1$}
\altaffiliation[Present address:~]{Department of Physics, Ritsumeikan University, Kusatsu, Shiga, 525-8577, Japan}
\author{Kiyoshi Kanazawa$^2$}
\altaffiliation[Present address:~]{Advanced Data Analysis and Modeling Unit, Institute of Innovative Research, Tokyo Institute of Technology, Midori-ku, Yokohama 226-8503, Japan}
\author{Hisao Hayakawa$^1$}
\affiliation{
	$^1$Yukawa Institute for Theoretical Physics, Kyoto University, Kitashirakawa-oiwake cho, Sakyo-ku, Kyoto 606-8502, Japan\\
	$^2$Department of Computational Intelligence and Systems Science, Interdisciplinary Graduate School of Science and Engineering, Tokyo Institute of Technology, 4259-G3-52 Nagatsuta-cho, Midori-ku, Yokohama 226-8503, Japan}

\begin{abstract}
This study numerically and analytically investigates the dynamics of a rotor under viscous or dry friction as a non-equilibrium probe of a granular gas.  
In order to demonstrate the role of the rotor as a probe for a non-equilibrium bath, the molecular dynamics (MD) simulation of the rotor is performed under viscous or dry friction surrounded by a steady granular gas under gravity. 
A one-to-one map between the velocity distribution function (VDF) of the granular gas and the angular distribution function for the rotor is theoretically derived. The MD simulation demonstrates that the one-to-one map accurately infers the local VDF of the granular gas from the angular VDF of the rotor, and vice versa. 
\end{abstract}
\maketitle

\section{Introduction}

Granular materials are ubiquitous in our daily life and are extensively studied mainly in various engineering fields such as soil mechanics, geology, powder technology, and civil engineering \cite{eng1,eng2}. They are also studied in the area of statistical physics \cite{nagel, aranson, poschel, brilliantov} because they exhibit various interesting phenomena, such as jamming transition and inhomogeneous cluster formation \cite{nagel, aranson, poschel, brilliantov}. These phenomena originate from the dissipative nature of granular particles during inelastic collisions. Granular gas is one of the simplest setups to theoretically understand the essence of granular materials. Indeed, the kinetic theory is applicable to granular gases when the density is not extremely high, where their dissipative nature appears as the non-Gaussian velocity distribution function (VDF) \cite{olafsen1,olafsen2,kudroli,kawarada,goldshtein_1995,esipov_1997,van_noije_1998,brilliantov_2000,dubey_2013}. 
Therefore, high-order cumulants, such as skewness and kurtosis, are expected to play important roles in understanding the characteristic behavior of granular materials in addition to the second-order cumulants (i.e., the granular temperature) \cite{hcs, micro_g, gdrmidi,forterre,cheng2,jet1,jet2,jet3, jam1,jam2,jenkins_savage_1983,jenkins_richman_1985,lutsko,garzo_dufty, van_noije_1998, hayakawa_2013, suzuki_2015}.

A typical method to measure VDFs includes the direct tracking of the motion of grains. This method is widely used for quasi-two dimensional systems. However, there are technical difficulties for three-dimensional systems because most of grains in the bulk of the system are invisible from the exteriors of containers. Although such difficulties have been overcome using magnetic resonance imaging \cite{mri} or fluorescent interstitial fluid \cite{imaging}, the former method is not easily accessible and the latter method induces additional rheological effects via the interstitial fluid on the granular flow. 

Another experimental method involves the indirect measurement of granular velocity fluctuation via probes. For example, a rotor can be placed into a granular gas as a probe to measure the angular velocity fluctuation of the rotor  \cite{brey, saraccino,devaraj,gnoli1,gnoli2,gnoli3,naert1,naert2,loreto_meer_2016}. The VDF of the granular gas can be inferred from that of the rotor. It is important to note that this method works well for systems in thermal equilibrium according to the fluctuation-dissipation relation. Recently, the relevance of this type of indirect method has been demonstrated for spatially homogeneous and isotropic granular gases through the analysis of the non-Gaussian Langevin equation in Refs \cite{kssh1,kssh2} on the basis of the Boltzmann-Lorentz equation \cite{talbot, cleuren,kampen,gardiner,smolchowski,chapman,kolmogorov}. 
However, the model used in this case is not sufficiently realistic because inhomogeneity and anisotropy exist in real granular gases such as vertically vibrated granular systems under gravity \cite{kudroli}. This implies that a more realistic formulation is necessary for the experimental measurement of high-order cumulants by observing the rotor dynamics. 

In this study, the molecular dynamics (MD) simulation of a realistic granular rotor is performed to demonstrate the role of a rotor as a probe to measure the VDFs of vibrating granular beds \cite{devaraj,gnoli1,gnoli2,gnoli3,naert1,naert2,loreto_meer_2016}.
An event driven MD simulation of a rotating rotor is performed around a fixed axis in a vertically vibrated granular gas under gravity in accordance with a method used in a previous study Ref. \cite{poschel}. 
The dynamics of the rotor in vibrating granular beds is analyzed to derive the relationship between the angular VDF of the rotor under viscous or dry friction and the VDF of the granular gas \cite{kssh1,kssh2}. 
In this study, it is also demonstrated that the formulas can be applied in the MD simulation.
The formulas can be used to infer the VDF of a gas with velocities that cannot be directly measured. Furthermore, it is demonstrated that the formulas work to detect the dependence of the VDF of the granular gas on its position in the container. Hence, the results indicate that the granular rotor can be used as a local velocity probe for a realistic granular gas. 

The organization of this paper is as follows. The setup of the simulation used in this study is described in Sec.~\ref{setup_sec}. In Sec.~\ref{gsse_sec} the basic equations to derive the inverse formula for a cylindrically symmetric granular gas \cite{kssh1,kssh2} are examined. In Sec.~\ref{vis_sec} the formulas for the VDF of the gas and the angular VDF of a rotor are derived under viscous friction around a rotating axis.
Furthermore, the numerical validity of these formulas are verified and the details of the numerical implementation are described. Section~\ref{sec_pos} discusses the dependence of the viscous rotor on its position in the container. In Sec.~\ref{dry_sec} the angular VDF for a dry frictional rotor is examined using the numerical VDF of the granular gas under gravity. In Sec.~\ref{summary_sec} the conclusions of the study are presented with some remarks. In Appendix \ref{detail_rotor_MD} the details for the MD simulation for the rotors are explained. Appendix \ref{benchmark} discusses a benchmark test of the simulation and formulation wherein the rotor is examined under viscous friction in an elastic gas without gravity, in which the angular VDF for the rotor is analytically obtained. Appendices \ref{app_vis} and \ref{app_dry} discuss the detailed derivation of the analytic formulas for the rotors under viscous and dry friction, respectively. 

\section{Setup of the simulation}\label{setup_sec}

\begin{figure}[h]
\begin{center}
\includegraphics[scale = 0.4]{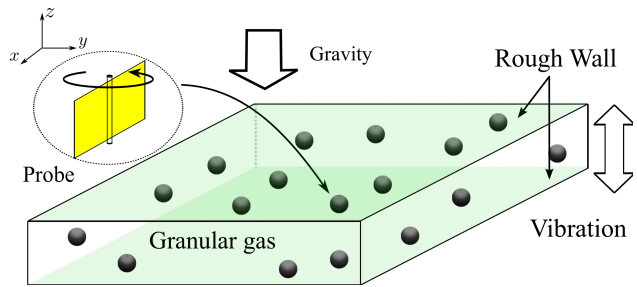}
\caption{(Color online) Schematics of the simulation. A rotating rotor (probe) is simulated in a vibrated granular gas under gravity.}
\label{setup}
\end{center}
\end{figure}

The schematics of the setup used in the study are illustrated in Fig.~\ref{setup}. 
The setup involves the preparation of $N = 100$ frictionless grains of diameter $d = 0.02\sqrt{A}$ and mass $m$ under gravity $g$ in a quasi two-dimensional container (area $A = L_{\rm box}^2$, height $H_{\rm box} = 0.1L_{\rm box}$). 
The VDF of inelastic grains under gravity and vibration is different from the Gaussian. The restitution coefficient $e_{\rm g} = 0.71$ is adopted for inelastic rigid grains, and it corresponds to the effective restitution coefficient for low density polyethylene~\cite{list_COR,list_COR1}. The restitution coefficient between grains and the side wall $e_{\rm w}$ is selected such that it is identical to that for collisions between the grains $(e_{\rm w} = e_{\rm g} = 0.71)$. The parameters in the simulation are summarized in Table \ref{table_setup}. Both the rotational motion of the grains and the tangential contact force between the grains are not considered in the event-driven MD simulation, because the effect of the tangential friction of spherical grains can be absorbed into the effective normal restitution coefficient if the duration (contact) time of the grains is negligible \cite{jenkins_zhang_2002, yoon_jenkins_2005,saitoh_hayakawa_2007,jenkins_2010}. It should be noted that the extension of the setup is straightforward for dense frictional grains in three-dimensional systems, where the effects of the rotations and the tangential frictions are not negligible. Appendix \ref{detail_rotor_MD} is referred to for the details of the event driven simulation.

\begin{figure}[h]
\begin{center}
\includegraphics[scale = 0.5]{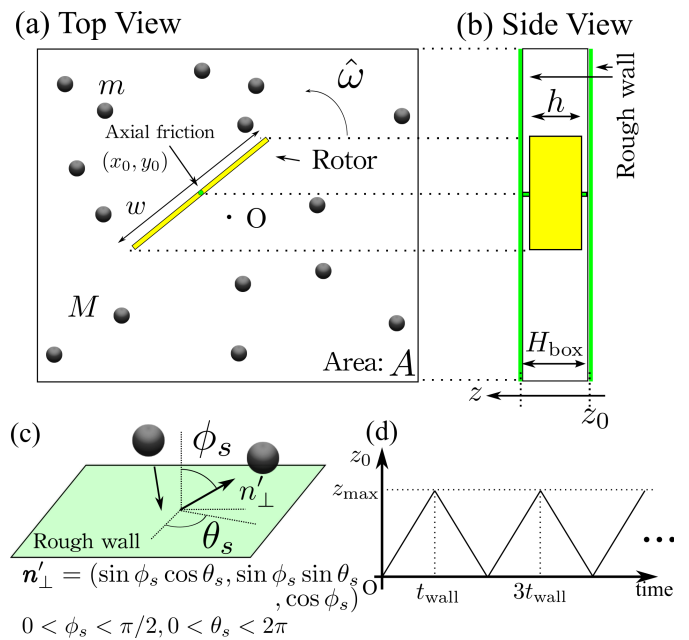}
\caption{(Color online) Schematics of (a) the top view and (b) the side view of the simulation. The rotor rotates around a fixed axis $(x_0, y_0)$, and the origin $O$ is the center of the container. Note that the size of the rotor is exaggerated in these figures. (c) The rough wall reflects the grain in a random direction. (d) The container is vibrated vertically in a piece-wise linear manner. The time evolution of the bottom of the container is shown.}
\label{setup1}
\end{center}
\end{figure}

\begin{table}[h]
\caption{Summary of the parameters for the simulation.}
\begin{tabular}{ccc}
\hline\hline
& symbols & values, definitions\\
\hline
mass ratio & $\epsilon$ & $m/M = 0.01$\\
rotor width & $w$ & $0.1L_{\rm box}$\\
grain diameter & $d$ & $0.02L_{\rm box}$\\
restitution coefficient & $e,e_{\rm g},e_{\rm w}$ &0.71\\
number of grains & $N$ &100\\
box height & $H_{\rm box}$ &$0.1L_{\rm box}$\\
rotor height& $h$ & $H_{\rm box}-d$\\
number density &$\rho$& $N/h(L_{\rm box}-d)^2$\\
vibration amplitude &$z_{\rm max}/2$& $0.01L_{\rm box}$\\
vibration period &2$t_{\rm wall}$& $\sqrt{2z_{\rm max}/g}$\\
vibration velocity &$v_0$& $z_{\rm max}/t_{\rm wall}$\\
observation radius &$r_{\rm obs}$& $2w$\\
\hline\hline
\end{tabular}
\label{table_setup}
\end{table}

\begin{figure*}[t]
\begin{center}
\includegraphics[scale = 1.3]{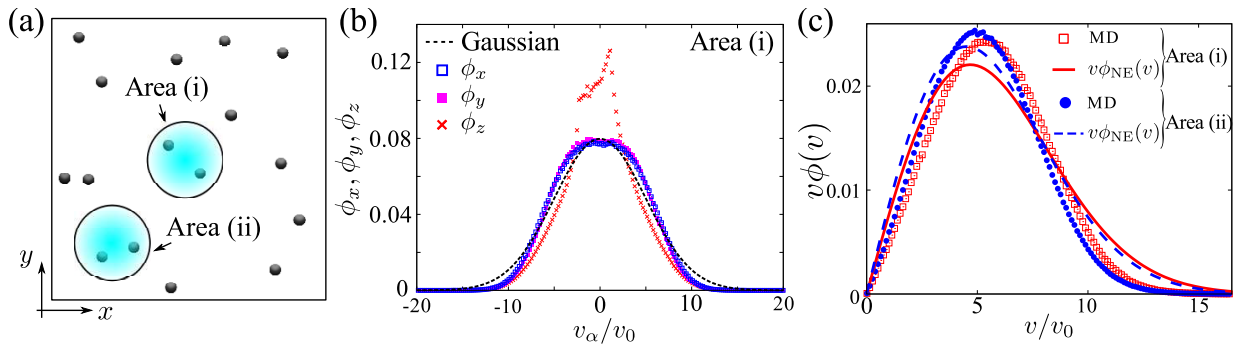}
\caption{(Color online)  (a): The area (i) is the cylindrical area with a radius $r_{\rm obs}$ around $x = 0, y = 0$, and the area (ii) is the area around $x = -L_{\rm box}/4, y = -L_{\rm box}/4$. (b): VDFs for $\alpha = x,y,z$ directions are shown. The data are observed in the area (i). VDFs in the horizontal direction are different from Gaussian (dotted line). The VDF in $z$ direction is asymmetric due to gravity. (c): The VDFs for $v = \sqrt{v_x ^2 + v_y ^2}$ in the areas (i) and (ii) are shown as open squares and filled circles, respectively. It is noted that the VDF for area (i) differs from that in the area (ii) because of the boundary effects. The observed VDFs cannot be fitted by the theoretical VDFs in Ref. \cite{van_noije_1998} represented by the solid line and the dashed line for the areas (i) and (ii), respectively.}
\label{vdf_area12}
\end{center}
\end{figure*}

The origin of the system in the laboratory frame $(x,y,z) = (0,0,0)$ is selected as the bottom center of the container at $t = 0$. A thin rotor of mass $M$ rotating around the fixed axis $(x, y) = (x_0, y_0)$ is introduced under the frictional torque $\hat{N}_{\rm fri}(\hat{\omega})$ with a width $w = 0.1L_{\rm box}$ and a height $h = H_{\rm box} - d$ (Fig.~\ref{setup1} (a) and (b)). The restitution coefficient $e$ between the rotor and grains is introduced and $e = e_{\rm w} = e_{\rm g} = 0.71$ is adopted. The moment of inertia of the rotor can be expressed as $I \equiv Mw^2 /12$. The density of the granular gas is $\rho = N/h(L_{\rm box} - d)^2$, where the volume fraction is given by $\pi d^3 \rho/6 \simeq 0.00545$. The rotor is assumed to be sufficiently massive, i.e., the mass ratio $\epsilon$ of the mass of the grain $m$ to that of the rotor $M$ can be obtained by $\epsilon \equiv m/M = 0.01 \ll 1$. The local VDF of the granular gas is measured near the rotating axis in the region of $r = \sqrt{(x-x_0)^2 + (y-y_0)^2}< r_{\rm obs} \equiv 2w$ and $z_0<z<z_0 + H_{\rm box}$.

Rough walls are introduced both on the top and the bottom of the container to distribute the energy in the horizontal direction (See Fig.~\ref{setup1} (b)). When a grain collides against the rough wall, the post collisional direction ${\bm n}' _{\perp}$ is randomized with the kinetic energy conserved during the collision. The scattered angles $(\theta_s, \phi_s)$ are selected from uniform random variables in $0\leq\phi_s\leq\pi/2, 0\leq\theta_s\leq2\pi$ (See Fig.~\ref{setup1} (c)). 
It is noted that the probability density per unit solid angle for small $\phi_s$ exceeds that for large $\phi_s$, while the probability density per unit solid angle for the horizontal direction $\theta_s$ is uniform. The rough walls introduced in this study corresponds to the walls where sandpapers are glued \cite{sandpaper}. Energy is injected into the granular gas by vertically vibrating the container in a piece-wise linear manner with a constant speed \cite{pise_wise_ref}. Figure \ref{setup1} (d) illustrates the time evolution of the bottom of the wall $z = z_0$. The direction of the container motion is changed by the interval $t_{\rm wall} = \sqrt{z_{\rm max}/2g}$. The amplitude is $z_{\rm max}/2 = 0.01L_{\rm box}$ and the speed of the box is given by $v_0 \equiv z_{\rm max}/t_{\rm wall}$. 
Note that the rough wall is different from the thermal wall, where the magnitude of velocity is randomly selected from the Maxwell distribution function \cite{maxwell1,maxwell2}. See Appendix \ref{benchmark} for the simulation of grains associated with a thermal wall.

We here show that the VDF of the gas is almost cylindrically symmetric. The VDFs are observed in two regions, namely in the areas (i) and (ii). Here, the center of the area (i) is $(x, y) = (0, 0)$, and the center of the area (ii) is $(x, y) = (-L_{\rm box}/4, -L_{\rm box}/4)$ (see Fig.~\ref{vdf_area12} (a)). The VDFs $\phi_{\alpha}$ for $\alpha = x,y,z$ directions are shown in Fig.~\ref{vdf_area12} (b), where the data are obtained in the area (i). It is noted that the VDFs for the horizontal direction also deviates from the Gaussian (dotted line) $\exp(-c_{\alpha} ^2/2)/\sqrt{2\pi}$ with $c_{\alpha} = v_{\alpha}/v_0$. The VDF for $v_z$ is irrelevant for the analysis because the rotor rotates around the vertical ($z$) axis and cannot detect the velocity in the $z$-direction. 
Thus, the inverse formula is formulated solely for horizontal VDFs on the basis of the Boltzmann-Lorentz equation. In Fig.~\ref{vdf_area12} (c), the numerical data of the VDFs of grains for $v \equiv \sqrt{v_x ^2 + v_y ^2}$ are shown for both areas (i) and (ii). It should be noted that the VDF in area (i) (open squares) differs from that in the area (ii) (filled circles) because of the boundary effect. In this study, only area (i) is considered in Secs. \ref{vis_sec} and \ref{dry_sec}, while both areas (i) and (ii) are discussed in Sec.~\ref{sec_pos}.

The obtained VDFs are compared with the theoretical VDFs for granular gases activated by a white noise thermostat \cite{van_noije_1998}, which is phenomenologically used for the analysis of vibrating granular gases \cite{micro_g}. Note that the observed VDFs can not be fitted by that in Ref. \cite{van_noije_1998}, which is expressed as $\phi_{\rm NE}(v) = (1 + a_2 S_2(v^2/v_{\rm th}^2))\phi_{\rm G}(v/v_{\rm th})/v_{\rm th}^2$ with $\phi_{\rm G}(c) \equiv \exp(-c^2/2)/2\pi$, $a_2 = 16(1-e_{\rm g})(1-2e_{\rm g}^2)/\{185 -153e_{\rm g}+ 30(1-e_{\rm g})e_{\rm g} ^2\}$, and $S_2(c) = c^2/2 - 2c + 1$. Here, $v_{\rm th}$ is a fitting parameter used in the setup. The fitting results are shown in Fig.~\ref{vdf_area12} (c) and $v_{\rm th}/v_0 = 4.67794$ and $v_{\rm th}/v_0 = 4.5033$ for areas (i) (chain line) and (ii) (dashed line), respectively.

\section{Basic equation}\label{gsse_sec}
The basic equation for cylindrically symmetric granular gases is described.  
We only consider the two-dimensional VDF $\phi = \phi(v_x, v_y)$ for the grains to calculate the angular VDF for the rotor. The time evolution of the probability distribution function (PDF) of the angular velocity of the rotor $P=P(\omega,t)$ can be described by the Boltzmann-Lorentz equation \cite{cleuren, talbot,gardiner,kampen,smolchowski,chapman,kolmogorov} as follows:
\begin{widetext}
\begin{eqnarray}
\frac{\partial P}{\partial t} + \left\{\frac{\partial}{\partial \omega}N_{\rm fri}P\right\} &=& \int_{-\infty} ^{\infty} dy \left\{W_{\epsilon}(\omega-y;y)P(\omega -y,t)-W_{\epsilon}(\omega;y)P(\omega,t)\right\}\label{bl_eq},\\
\epsilon W_{\epsilon}(\omega;y) &\equiv& \rho h \int_0 ^{2w} d{\sigma}\int_{-\infty} ^{\infty}dv_xdv_y\phi(v_x,v_y)\Theta({V}_n({\sigma}) -{v}_n)|{V}_n({\sigma}) -{v}_n|\delta\left(\frac{y}{\epsilon} -\Delta\bar{\omega}\right).\label{w_rate}
\end{eqnarray}
\end{widetext}
Here, the following expression is introduced ${\bm V}({\sigma}) \equiv \omega{\bm e}_z\times{\bm r}({\sigma}), \Delta\bar{\omega} \equiv g({\sigma})(1+e){(V_n - v_n)}/\{R_I(1+\epsilon g^2({\sigma}))\}, g({\sigma}) \equiv r_t({\sigma})/R_I, {\bm t}({\sigma}) \equiv {\bm e}_z\times{\bm n}({\sigma})$, and $R_I \equiv \sqrt{I/M}$, where ${\bm n}$ and ${\bm t}$ denote normal and tangential unit vectors on the surface of the rotor, respectively. Correspondingly, the variables with the subscripts $n$ and $t$ denote the normal and the tangential components of the vectors, respectively. The unit vector in $z$ direction is expressed as ${\bm e}_z$. It should be noted that this set of equations is widely used in systems such as granular gases activated by a white noise thermostat \cite{evans_morris, puglisi_1998,van_noije_1999,gradenigo_2011,khalil_2014,van_noije_1998}. Additionally, ${\sigma}$ is also introduced as a coordinate variable along the surface of the rotor, running over $0<{\sigma}<2w$ \cite{cleuren}.
According to Refs.~\cite{kssh1,kssh2}, Equation (\ref{bl_eq}) is reduced to a Langevin equation for the angular velocity $\Omega\equiv \omega/\epsilon$ driven by the state-independent non-Gaussian noise in the massive rotor limit $\epsilon\to 0$ when the axial friction is sufficiently strong. 
The following section examines the steady distribution function denoted by $P_{\rm ss}(\Omega)\equiv\lim_{t\to\infty}{\mathcal P}(\Omega, t)$ with ${\mathcal P}(\Omega,t) \equiv \epsilon P(\epsilon\Omega, t)$ for the two types of axial frictions, namely the viscous friction $N_{\rm fri} = - \gamma\omega$ and the dry friction $N_{\rm fri} = -\Delta{\rm sgn}(\omega)$, in which the signature function ${\rm sgn}(x)\equiv x/|x|$ with ${\rm sgn}(0) = 0$ and the friction coefficients $\gamma$ and $\Delta$ \cite{persson,friction_h,genne,friction_review,sano_hayakawa} are introduced.

\begin{figure*}[t]
\begin{center}
\includegraphics[scale = 1.3]{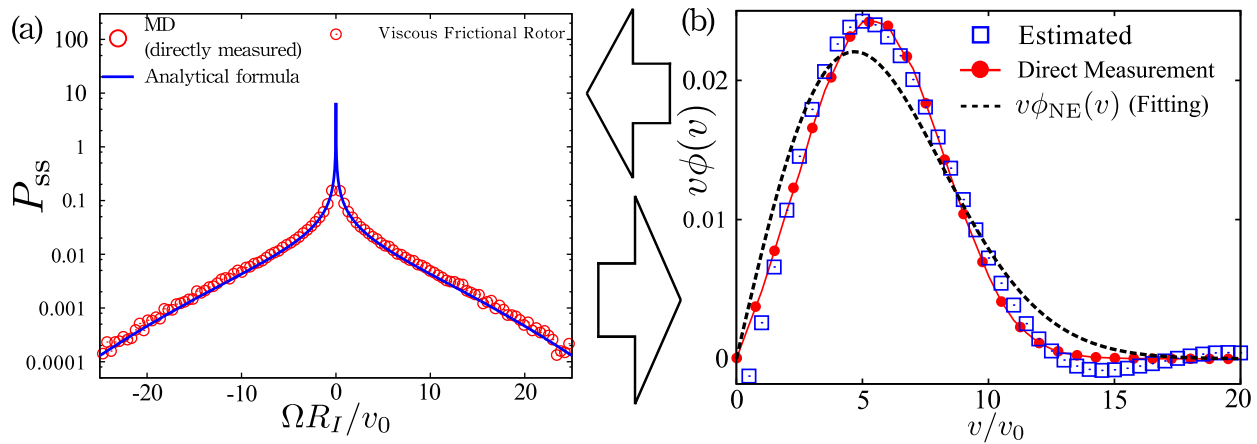}
\caption{(Color online) The demonstration of the applicability of the formulas (\ref{pss_analytical})-(\ref{g_k}) under viscous friction on (a) the forward and (b) the inverse estimation problems. The Fourier transform of Eq.~(\ref{pss_analytical}) is shown as the solid line (a), which corresponds to the directly measured MD data. (b) The formulas (\ref{phi_solved}) represented by open squares are compared with the numerical data for the VDF $\phi(v)$ near the rotor (filled circles). The VDF of the granular gas can be successfully estimated by only observing the angular VDF of the rotor. Note that a numerical error exists for large $v/v_0$.}
\label{inverse}
\end{center}
\end{figure*}

\section{Granular Rotor under Viscous Friction}\label{vis_sec}
In this section, the role of the rotor under viscous friction $N_{\rm fri} = - \gamma\omega$ is examined in terms of a probe of the granular gas. In Sec.~\ref{analytical_sec_vis}, the forward and inverse formulas connecting the granular VDF with the rotor PDF are analytically derived. In Sec.~\ref{forward_vis_sec}, the validity of the forward formula is verified by estimating $P_{\rm ss}(\Omega)$ using the numerical data for $\phi(v)$. Next, in Sec.~\ref{inverse_sec}, the inverse problem is solved, i.e., the granular VDF $\phi(v)$ is derived from a given $P_{\rm ss}(\Omega)$, which enables the inference of the properties of granular gases from the motion of the probe, i.e., the rotor. Section~\ref{viscous_detail_sec} describes the detailed procedures and numerical techniques applied to derive the formulas.

\subsection{Analytic formulas for PDF  of the rotor}\label{analytical_sec_vis}

With the aid of Ref.~\cite{eliazar}, the steady angular VDF in the Fourier transform $\tilde{P}_{\rm ss}(s) \equiv \int_{-\infty} ^{\infty} d\Omega e^{is\Omega}P_{\rm ss}(\Omega)$ can be expressed as follows:
\begin{eqnarray}
\tilde{P}_{\rm ss}(s) &=& \exp\left[\int_0 ^s \frac{I}{\gamma s'}\Phi(s')ds'\right]\label{pss_analytical},
\end{eqnarray}
where the cumulant generating function $\Phi(s) \equiv\sum_{l=1}^{\infty} {{\mathcal K}_l}(is)^l/{l!} = \int_{-\infty} ^{\infty}{\mathcal W}({\mathcal Y}) (e^{i{\mathcal Y}s} -1)$ with ${\mathcal K}_l \equiv \int_{-\infty} ^{\infty} d{\mathcal Y} {\mathcal Y}^l{\mathcal W}({\mathcal Y})$ and the scaled transition rate ${\mathcal W}(\mathcal Y)$ can be represented by an integral transform of $\phi(v)$ for the cylindrically symmetric case as follows:
\begin{eqnarray}
\Phi(s) =-  \frac{2\rho hwv_0}{\tilde{s}^2\tilde{w}^2} \int_0 ^{\infty} d\tilde{v}\tilde{\phi}(\tilde{v})\left\{-(\tilde{w}\tilde{s}\tilde{v})\pi{\rm H}_0(\tilde{w}\tilde{s}\tilde{v}) \right.\nonumber\\ \left.+2(\tilde{w}\tilde{s}\tilde{v})^2\right\}\label{Psi_rotor}.
\end{eqnarray}
Here, dimensionless variables $\tilde{w} = (1+e)w/2R_I, \tilde{s} = sv_0/R_I, \tilde{v}= v/v_0$, and $\tilde{\phi}(\tilde{v}) = v_0^2 \phi(v_0 \tilde{v})$ are introduced. The Struve function ${\rm H}_{\nu}(x)$ with $\nu = 0$ defined by Eq.~(\ref{struve0_def}) \cite{abramowitz} is used. Appendix \ref{app_vis} provides the detailed derivation. It should be noted that Eq.~(\ref{Psi_rotor}) is valid for both the rotor under linear (viscous) friction and for the rotor under nonlinear (dry) friction. The following expression is obtained by substituting Eq.~(\ref{Psi_rotor}) into Eq.~(\ref{pss_analytical}):
\begin{eqnarray}
\frac{\tilde{\gamma}}{\pi k}\left\{k^3 \frac{d}{dk}{\rm ln}\tilde{P}_{\rm ss}\left(\frac{k}{\tilde{w}}\right)+Bk^2\right\} = \int_0 ^{\infty}d\tilde{v} \tilde{v}\tilde{\phi}(\tilde{v}){\rm H}_0(k\tilde{v}),\nonumber\\
\label{eq_last_main}
\end{eqnarray}
where $B = (2/\tilde{\gamma})\int_0 ^{\infty}d\tilde{v} \tilde{v}^2 \tilde{\phi}(\tilde{v})$, a dimensionless variable $k \equiv \tilde{w}\tilde{s}$, and a scaled friction coefficient $\tilde{\gamma} \equiv \gamma/(2\rho hwIv_0)$ are introduced. The integral on the right hand side of Eq.~(\ref{eq_last_main}) is known as the Struve transformation, and its inverse transformation is the $Y$ transform. These are types of the Bessel transformations \cite{titchmarsh}. The inverse estimation formula is obtained by introducing the Neumann function $N_{\nu}(x)$ with $\nu = 0$ \cite{abramowitz}:
\begin{eqnarray}
\tilde{\phi}(\tilde{v}) &=& \int_0 ^{\infty} G_{\rm vis}(k)N_0(k\tilde{v})kdk,\label{phi_solved}\\
G_{\rm vis}(k) &\equiv& \frac{\tilde{\gamma}}{\pi k}\left\{k^3 \frac{d}{dk}{\rm ln}\tilde{P}_{\rm ss}\left(\frac{k}{\tilde{w}}\right)+Bk^2\right\}.\label{g_k}
\end{eqnarray}
The VDF of the granular gas is determined from Eqs.~(\ref{phi_solved}) and (\ref{g_k}) by observing the rotor dynamics. This implies that the rotor is considered as a thermometer for the granular gas with the aid of the inverse formula Eqs.~(\ref{phi_solved}) and (\ref{g_k}). Note that the constant $B$ in Eq.~(\ref{g_k}) is numerically determined by the condition $\lim_{k\to\infty}G_{\rm vis}(k) = 0$ known as the Riemann-Lebesgue lemma. 

\subsection{Forward problem for viscous rotor}\label{forward_vis_sec}
Prior to considering the inverse problem, the forward problem, i.e., the determination of the PDF of the rotor from the VDF of the granular gas is discussed. The validity of the formulas (\ref{pss_analytical}) and (\ref{Psi_rotor}) are verified using the numerical VDF $\phi(v)$ of the granular gas under gravity. In the following numerical simulation, $\gamma/Mv_0z_{\rm max} = 5.0$, which corresponds to $\tilde{\gamma} = 5.7624$. In Fig.~\ref{inverse} (a), $P_{\rm ss}(\Omega)$ is plotted by the solid line on the basis of Eqs.~(\ref{pss_analytical}) and (\ref{Psi_rotor}) as well as the numerical histogram for $\phi(v)$ and the results of the MD simulation (circles). In general, the theory agrees with the MD simulation except for the point near $\Omega = 0$, where the numerical error can be reduced if smaller bin-width is used for $\phi(v)$. The bin-width of the numerical histogram is adopted for $\phi(v)$ as $2.5\times 10^{-2}v_0$. Section~\ref{vis_detail_for} provides the detailed procedure to obtain the solid line in Fig \ref{inverse} (a).

\subsection{Inverse problem for granular gas}\label{inverse_sec}
This subsection discusses that the VDF of the granular gas can be inferred only through the numerically obtained steady distribution of the angular velocity of a rotor under the viscous friction by using the formula Eqs.~(\ref{phi_solved}) and (\ref{g_k}). 

The result for the inverse formula in Eqs.~(\ref{phi_solved}) and (\ref{g_k}) is shown in Fig.~\ref{inverse} (b), where the parameter $B$ is estimated as $B=0.365556$ and we use the bin-width $4.33011\times 10^{-4} v_0/R_I$ for $P_{\rm ss}$. The formula (\ref{phi_solved}) represented by open squares correctly predicts the numerical VDF $\phi(v)$ near the rotor (filled circles), while the theoretical VDF $v\phi_{\rm NE}$ for the white noise thermostat fails to fit the data. The bin-width for $P_{\rm ss}$ is considered as $4.33011\times 10^{-4} v_0/R_I$. Although numerical oscillations exists for large $v/v_0$, the estimation for the granular gas on the basis of Eq.~(\ref{phi_solved}) corresponds well to the directly measured VDF of granular particles in the MD simulation. This implies that the inverse formula (\ref{phi_solved}) supplemented by Eq.~(\ref{g_k}) enables the use of the rotor as a non-equilibrium thermometer. Sec.~\ref{vis_detail_inv} provides the details of the numerical implementation.

\subsection{Numerical implementation for the formulas Eqs.~(\ref{pss_analytical})- (\ref{g_k})}\label{viscous_detail_sec}
\begin{figure}[h]
\begin{center}
\includegraphics[scale = 0.42]{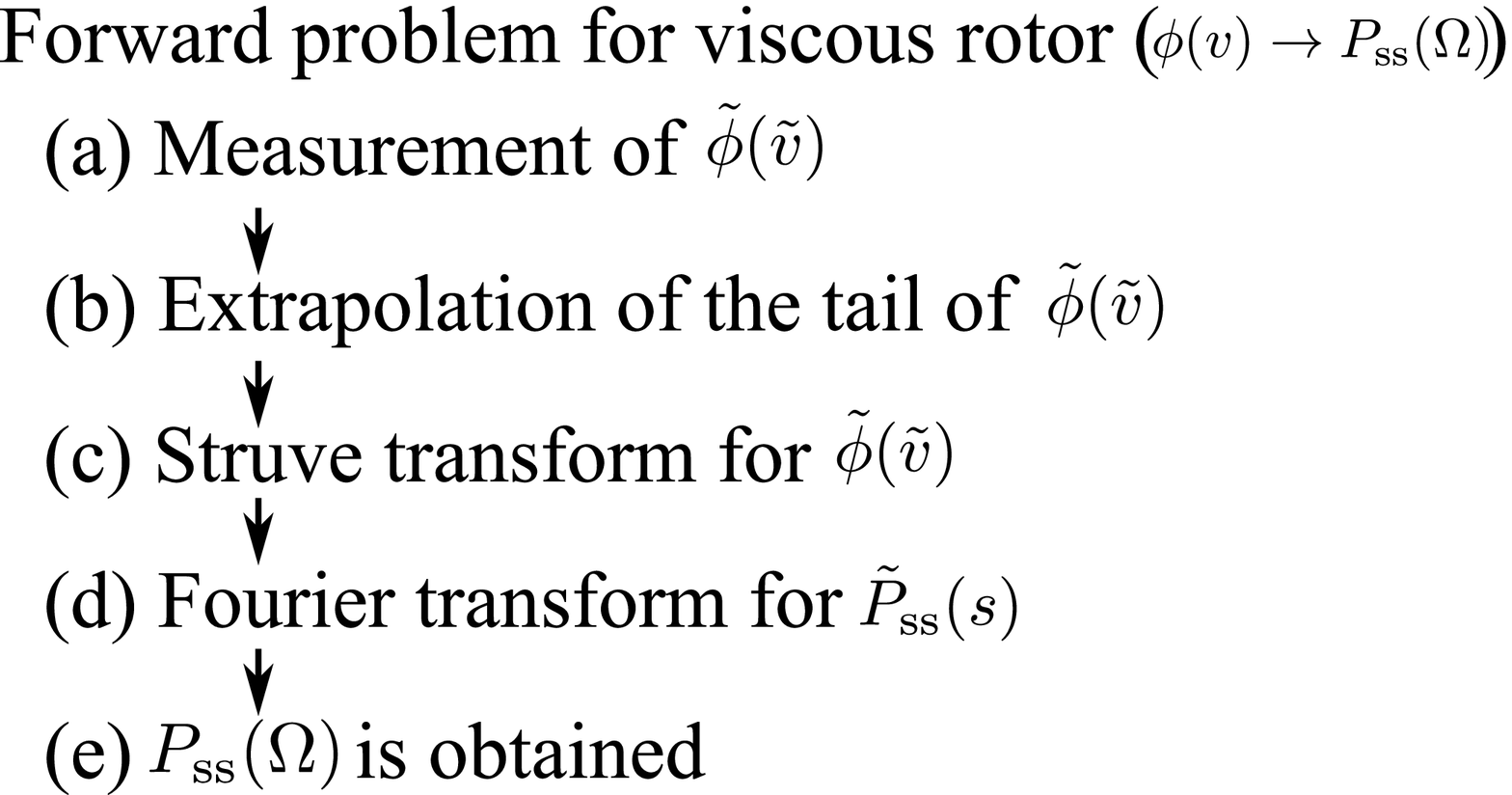}
\caption{(Color online) Outline of the forward problem for the viscous rotor.}
\label{proc_forward_vis}
\end{center}
\end{figure}

This subsection describes the technical details of the numerical implementation for the formulas Eqs.~(\ref{pss_analytical})-(\ref{g_k}). The description is useful for both experimentalists and theoreticians interested in the details of the method. However, this subsection can be skipped if readers are not interested in the technical details.

\subsubsection{Forward problem}\label{vis_detail_for}
The detailed numerical technique is described to obtain $P_{\rm ss}(\Omega)$ using Eqs.~(\ref{pss_analytical}) and (\ref{Psi_rotor}). Figure~\ref{proc_forward_vis} can be referred to for the outline of the procedure. 
The numerical technique contains the following five steps (a)-(e):

(a) The VDF of the granular gas $\tilde{\phi}(\tilde{v})$ around the rotor is measured. 

(b) The data of $\tilde{v}\tilde{\phi}(\tilde{v})$ are extrapolated for the tail ($\tilde{v}\to\infty$). The data are fitted in the range $\tilde{v}_-<\tilde{v}<\tilde{v}_+$ by a function $b_1 \exp(-b_2 \tilde{v})$ with fitting parameters $b_1$ and $b_2$, and the data are then extrapolated to the range $\tilde{v}_-<\tilde{v} < \tilde{v}_{\rm end} = 20\tilde{v}_{-}$. Here, $\tilde{v}_-, \tilde{v}_+$, and $\tilde{v}_{\rm end}$ denote the end points for the fitting ranges. It is verified that the following results are invariant even if the Gaussian is used as a fitting function instead.

(c) The Struve transform is applied for $\tilde{\phi}(\tilde{v})$ according to Eqs.~(\ref{pss_analytical}) and (\ref{Psi_rotor}) to obtain $\tilde{P}_{\rm ss}(s)$. 

(d) The Fourier transform for $\tilde{P}_{\rm ss}(s)$ is used to obtain $P_{\rm ss}(\Omega)$. Note that $P_{\rm ss} (\Omega)$ had a sharp peak around $\Omega=0$, which is serious for the numerical Fourier transformation in terms of convergence. This problem is solved by using the double exponential formula, which is a numerical technique for singular functions \cite{double_exponential}. 

(e) $P_{\rm ss}(\Omega)$ is obtained. The fitting parameters adopted here are listed in Table \ref{num_table_forward_vis}.

\begin{figure}[h]
\begin{center}
\includegraphics[scale = 0.4]{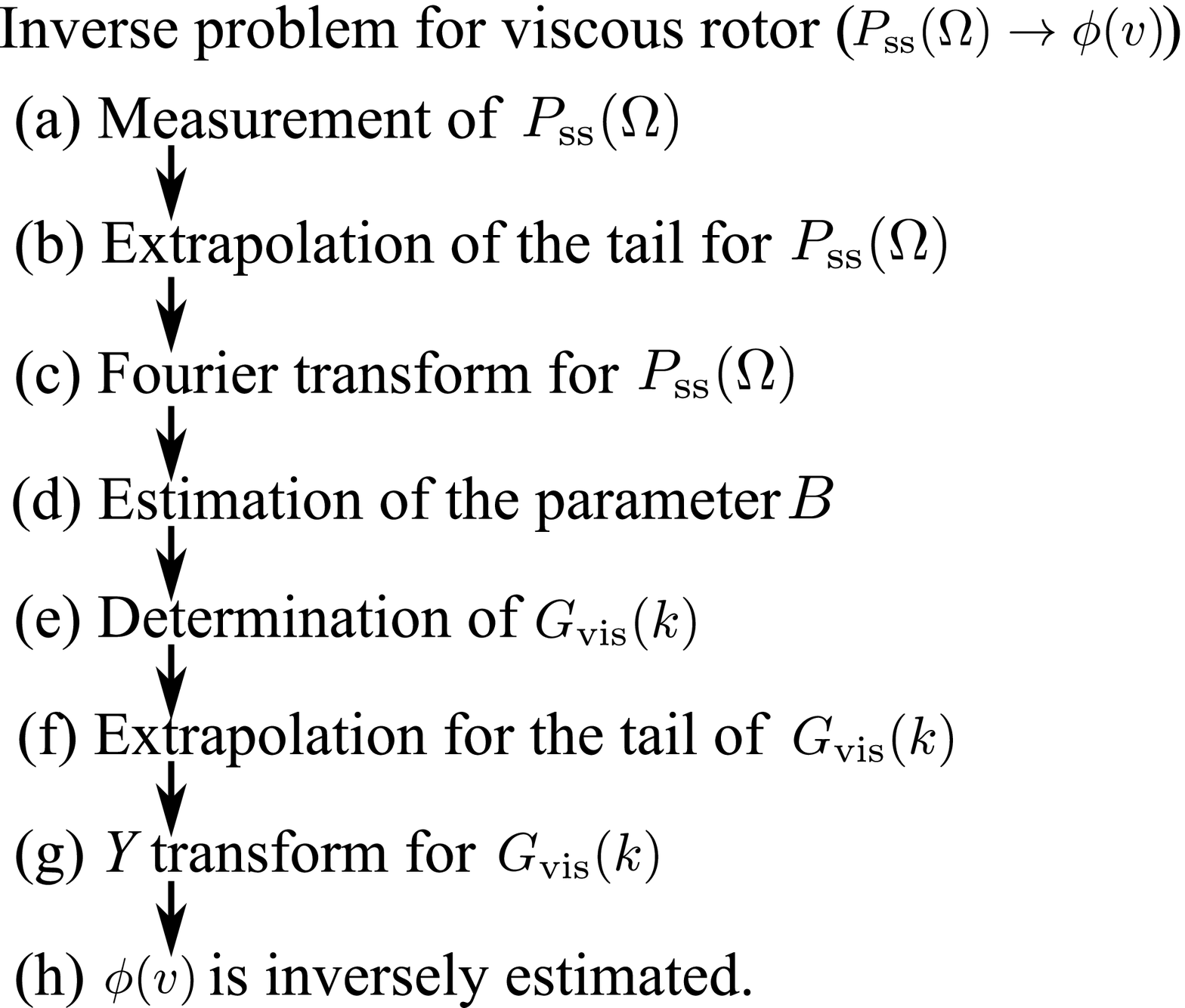}
\caption{(Color online) Outline of the inverse problem for the viscous rotor.}
\label{proc_inv_vis}
\end{center}
\end{figure}

\begin{table}[h]
\caption{Summary of numerical values in the forward problem for the viscous rotor.}
\begin{tabular}{cc}
\hline\hline
symbols & values\\
\hline
$\tilde{v}_-$ &  15.0\\
$\tilde{v}_+$ & $25.0$\\
$\tilde{v}_{\rm end}$ & $300.0$\\
$b_1$ & $1182.192$\\
$b_2$ & $1.128878$\\
\hline\hline
\end{tabular}
\label{num_table_forward_vis}
\end{table}

\subsubsection{Inverse problem}\label{vis_detail_inv}
It is necessary to discuss the detailed numerical technique to obtain $\phi(v)$ from $P_{\rm ss}(\Omega)$ on the basis of Eqs.~(\ref{phi_solved}) and (\ref{g_k}). Fig.~\ref{proc_inv_vis} shows the outline of the procedure. The procedure includes the following eight steps (a)-(h) to obtain $\tilde{\phi}$ from numerical $P_{\rm ss}(\Omega)$.

(a) The angular VDF of the rotor $P_{\rm ss}(\Omega)$ is measured.

(b) This is followed by the extrapolation of $P_{\rm ss}(\Omega)$ for the tail ($\Omega\to\infty$). Furthermore, $P_{\rm ss}(\Omega)$ is fitted in the range $\Omega_-<\Omega<\Omega_+$ by a fitting function $b_1 ' \exp(-b_2 ' \Omega)$ and $P_{\rm ss}(\Omega)$ is extrapolated for the range $\Omega_-<\Omega < \Omega_{\rm end}$ with the  cut off $\Omega_{\rm end}$. Here, $\Omega_-$ and $\Omega_+$ denote the end points for fitting and $b_1 '$ and $b_2 '$ denote the fitting parameters. It is also verified that the following results are invariant even if a Gaussian fitting function is used.

(c) The Fourier transform for extrapolated $P_{\rm ss}(\Omega)$ is performed to obtain $\tilde{P}_{\rm ss}(s)$. The double exponential formula \cite{double_exponential} is similarly applied to the forward problem.

(d) The parameter $B$ is estimated and the data of $k^3 d \log \tilde{P}/dk$ for $k\to \infty$ are fitted by using the quadratic function $c' - Bk^2$ in the region $k_c ^-<k<k_c ^+$. Here, $k_c ^-$ and $k_c ^+$ denote the end points for fitting, and the fitting parameter $c'$ is introduced.

(e) Furthermore, $G_{\rm vis}(k)$ is calculated using Eq.~(\ref{g_k}). 

(f) Numerical $G_{\rm vis}(k)$ is fitted by a fitting function $c/k + d/k^3$ with fitting parameters $c$ and $d$ in the $k\to\infty$ limit. Moreover, $G_{\rm vis}(k)$ is fitted in the region $k_e ^-<k<k_e ^+$, and $G_{\rm vis}(k)$ is extrapolated for $k_{\rm cut} '<k<k_{\rm end} '$, where the cutoffs satisfy $k_e ^-<k_{\rm cut} '<k_e ^+$. The end points of the fitting ranges: $k_e ^-,k_e ^+$, $k_{\rm cut}'$, and $k_{\rm end}$ are introduced to obtain the asymptotic behavior of $G_{\rm vis}(k)$ for $k\to\infty$.

(g) The $Y$ transform (\ref{phi_solved}) is applied for the numerically obtained $G_{\rm vis}(k)$.

(h) $\phi(v)$ is inversely estimated. 
The fitting ranges and parameters introduced here are summarized in Table \ref{table_inverse_vis1}.

\begin{table}[h]
\caption{Summary of numerical values in the inverse problem for the viscous rotor (Area (i)).}
\begin{tabular}{cc}
\hline\hline
symbols & values\\
\hline
$\Omega_-$ & $20.0 v_0/R_I$\\
$\Omega_+$ & $25.0 v_0/R_I$\\
$\Omega_{\rm end}$ & $5.0 \Omega_-$\\
$k_c ^-$ & 1.0\\
$k_c ^+$ & 1.6\\
$k_e ^-$ & 0.5\\
$k_e ^+$ & 0.7\\
$d$ & 0.004827312\\
$k_{\rm cut} '$& 0.6\\
$k_{\rm end} '$ & $1.1936266\times 10^{4}$\\
$b_1 '$ & $0.0876136 R_I/v_0$\\
$b_2 '$ & $0.253986 R_I/v_0$\\
$c'$ & 0.0162884\\
$c$ & 0.01885404\\
\hline\hline
\end{tabular}
\label{table_inverse_vis1}
\end{table}

\begin{figure}[t]
\begin{center}
\includegraphics[scale = 0.7]{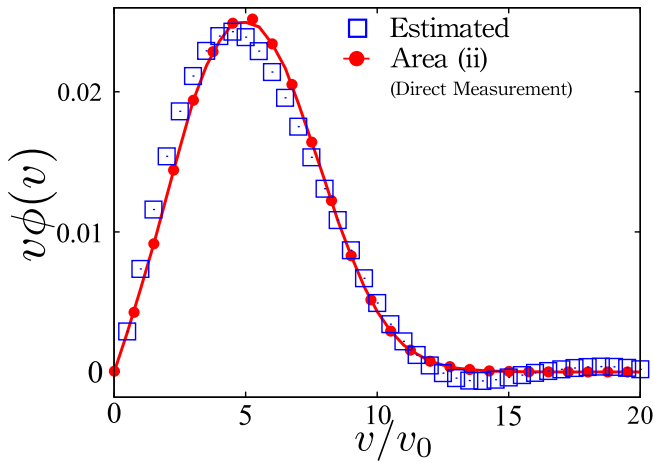}
\caption{(Color online) The demonstration of the inverse formula for the local VDF in the area (ii). Although the accuracy is lower than that in Fig.~\ref{inverse}, the estimated data (squares) correctly predict the direct measurement data (circles). }
\label{inverse_diff_pos}
\end{center}
\end{figure}

\begin{figure*}[t]
\begin{center}
\includegraphics[scale = 1.35]{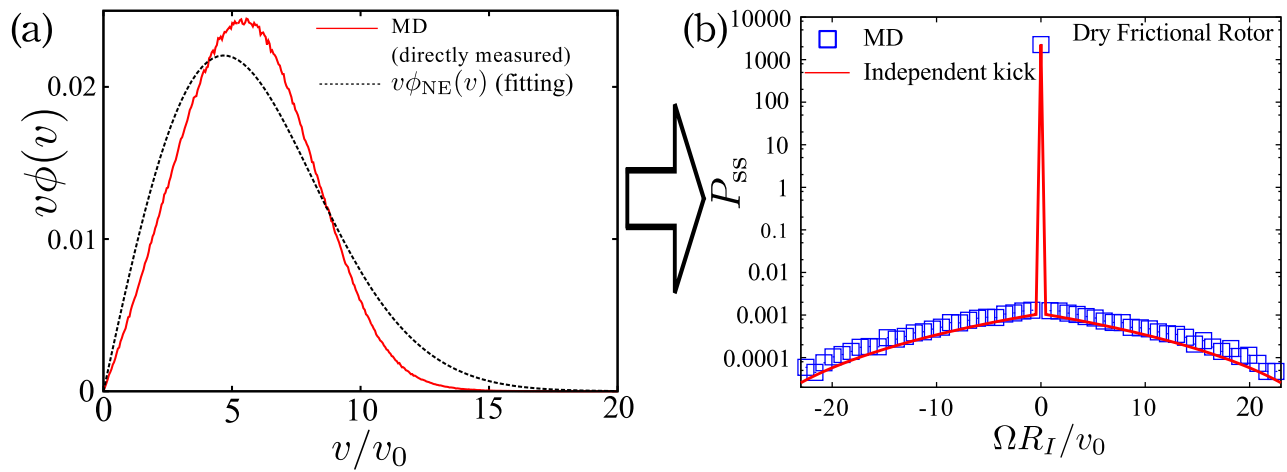}
\caption{(Color online) (a): Eq.~(\ref{sol_ik_bessel}) is adopted for the VDF of the gas shown as the solid line. VDF in Ref. \cite{van_noije_1998} is also shown. (b): $P_{\rm ss}(\Omega)$ of the dry frictional rotor is calculated according to Eq.~(\ref{sol_ik_bessel}). The squares and the solid line are histograms for MD and the Fourier transform of Eq.~(\ref{sol_ik_bessel}), respectively. The observed histogram in the MD simulation can be correctly predicted by the Fourier transform of Eq.~(\ref{sol_ik_bessel})} 
\label{pdf_dry_md}
\end{center}
\end{figure*}

\section{Position dependence of the rotor}\label{sec_pos}
This section discusses the utility of the rotor as a local non-equilibrium probe. In Sec.~\ref{inverse_sec} it is already demonstrated that the VDF of the surrounding gas in the area (i) can be inferred from the angular VDF of the viscous rotor $P_{\rm ss}$. However, as shown in Fig.~\ref{vdf_area12} (c), VDFs in areas (i) and (ii) have slight differences. It is important to discuss whether or not the local VDF in the area (ii) can be inferred from the motion of the rotor using Eqs.~(\ref{phi_solved}) and (\ref{g_k}).

In Fig.~\ref{inverse_diff_pos}, it is shown that the VDF in area (ii) can be also inferred from the angular VDF of the rotor. The estimated VDF and the directly measured VDF are represented by squares and circles, respectively. Here, the parameter $B$ is estimated as $B = 0.40278$. Numerical oscillations for large $v/v_0$ also exists. Although the accuracy of estimation is slightly lower than that in Fig.~\ref{inverse} (b), the estimated data are consistent with the directly observed VDF. The reason for the small discrepancy can be attributed to the violation of the cylindrical symmetry because of the boundary effect. The same procedure as Section~\ref{vis_detail_inv} is adopted, and the corresponding fitting ranges and parameters in this section are listed in Table \ref{table_inverse_vis2}.

\begin{table}[h]
\caption{Summary of numerical values in the inverse problem for the viscous rotor (Area (ii)).}
\begin{tabular}{cc}
\hline\hline
symbols & values\\
\hline
$\Omega_-$ & $20.0 v_0/R_I$\\
$\Omega_+$ & $25.0 v_0/R_I$\\
$\Omega_{\rm end}$ & $20.0\Omega_-$\\
$k_c ^-$ & 1.0\\
$k_c ^+$ & 1.6\\
$k_e ^-$ & 0.5\\
$k_e ^+$ & 0.7\\
$d$ & 0.004827312\\
$k_{\rm cut} '$& 0.6\\
$k_{\rm end} '$ & $1.1936266\times 10^{4}$\\
$b_1 '$ & $0.0927576 R_I/v_0$\\
$b_2 '$ & $0.262268 R_I/v_0$\\
$c'$ & 0.02641632\\
$c$ & 0.0339546\\
\hline\hline
\end{tabular}
\label{table_inverse_vis2}
\end{table}

\section{Granular Rotor under Dry Friction}\label{dry_sec}
A rotor under dry friction $N_{\rm fri} = -\Delta{\rm sgn}(\omega)$ is considered. Note that the real experimental rotors are influenced by dry friction \cite{gnoli1,gnoli2,gnoli3}. In Sec.~\ref{dry_ana_sec} we show the outline of the derivation of an analytic formula for the angular VDF of the rotor, and verify its validity in Sec.~\ref{dry_sim_sec} (See Appendix. \ref{app_dry} for the detailed derivation). Note that only the forward problem is examined by the perturbative method developed in Refs. \cite{kssh2, talbot}. 

\subsection{Analytic formulas for the PDF  of the rotor}\label{dry_ana_sec}
 The steady angular VDF $P_{\rm ss}(\Omega)$ is obtained perturbatively in terms of $1/\tilde{\Delta}'$ with $\tilde{\Delta}' \equiv {\Delta}(1+e)/(\epsilon I\rho hv_0 ^2 4{\pi}) \gg 1$ from Eq.~(\ref{bl_eq}) after taking the limit $\epsilon\to 0$. The first-order solution in terms of $1/\tilde{\Delta}'$ is known as the independent kick solution, which is originally introduced in Ref. \cite{talbot} and is systematically derived in Refs. \cite{kssh1,kssh2}. The independent kick solution is expressed as:
\begin{eqnarray}
\tilde{P}_{\rm ss}(s) &=& 1 + \frac{I}{\tilde{\Delta}}\int_{-\infty} ^{\infty}d{\mathcal Y} {\mathcal W}({\mathcal Y})\int_0 ^{\mathcal Y} \frac{d\Omega}{{\rm sgn}(\Omega)}\left[e^{is\Omega} -1\right]\nonumber\\ &&+ O(1/\tilde{\Delta}^{2})\label{sol_ik},
\end{eqnarray}
with the friction coefficient independent of $\epsilon$: $\tilde{\Delta} \equiv \Delta/\epsilon$.
Introducing the Bessel function $J_{\nu}(x)$ \cite{abramowitz}, $\tilde{P}_{\rm ss}$ of the rotor under the dry friction can be rewritten as:
\begin{eqnarray}
\tilde{P}_{\rm ss}\left(\frac{k}{\tilde{w}}\right) &=& 1 + \frac{\tilde{w}^2}{k^2}\frac{1}{\tilde{\Delta}'}\left[\frac{1}{2\pi}+ C_1k^2 -G_{\rm dry}(k)\right]\nonumber\\
&&+ O(1/\tilde{\Delta}^{'2}) \label{sol_ik_bessel},\\
G_{\rm dry}(k) &\equiv&\int_0 ^{\infty}d\tilde{v}\tilde{v}\tilde{\phi}(\tilde{v})J_0(k\tilde{v})\label{g_dry},
\end{eqnarray}
where the following coefficient is introduced:
\begin{equation}
C_l \equiv \frac{(-1)^l}{2^{2l} (l!)^2} \int_0 ^{\infty}d\tilde{v}\tilde{v}^{2l+1} \tilde{\phi}(\tilde{v})\label{c_n},
\end{equation}
for a positive integer $l = 1,2,\cdots$. Note that $\tilde{P}_{\rm ss}(s)$ can be separated into the $\delta$-type singular part denoted by $P_{\rm inf} \equiv \lim_{s\to\infty} \tilde{P}_{\rm ss}(s) =1 +C_1\tilde{w}^2/\tilde{\Delta}'$ and the smooth part denoted by $\tilde{P}_{\rm ss} ^{c}(s) \equiv  \tilde{P}_{\rm ss}(s) - P_{\rm inf}$. 
See Appendix \ref{app_dry} for the detailed derivation.

\subsection{Forward problem for dry frictional rotor}\label{dry_sim_sec}
This section examines the consistency between the theoretical results in  Eqs.~(\ref{sol_ik_bessel}) and (\ref{g_dry}) and the numerical results in the MD simulation under dry friction. For the MD simulation, we adopt $\Delta /Mv_0 ^2= 5.0$, which corresponds to $\tilde{\Delta}' = 784.13275$. The bin-width for $P_{\rm ss}$ is set as $4.33011\times 10^{-4} v_0/R_I$. As shown in Fig.~\ref{pdf_dry_md} (b), the numerical PDF $P_{\rm ss}(\Omega)$ in the MD simulation is correctly predicted from the numerical VDF of the granular gas $\phi(v)$ according to Eq.~(\ref{sol_ik_bessel}) (Fig.~\ref{pdf_dry_md} (a)).
It is also verified that the VDF of the granular gas does not depend on the details of the $\omega$ dependence of the frictional torque $N_{\rm fri}$.
It is noted that the VDF of the granular gas obtained in the MD simulation can not be fitted by that activated by the white noise thermostat \cite{van_noije_1998}. 
In Fig.~\ref{pdf_dry_md} (b), the squares and the solid line represent histograms for MD and the Fourier transform of Eq.~(\ref{sol_ik_bessel}), respectively. For a given VDF of the granular gas, the theoretical result agrees with the result of the MD simulation without introducing any fitting parameters. Note that there are small discrepancies between the theoretical result and the data for the large $\Omega R_I/v_0$, because the independent kick model is accurate only for small $\Omega R_I/v_0$. The detailed implementation of Eqs.~(\ref{sol_ik_bessel}) and (\ref{g_dry}) are provided in Sec.~\ref{dry_detail_sec}. 

We here discuss the difficulty for the inverse estimation problem under dry friction in the present theoretical analysis. Although Eq.~(\ref{sol_ik_bessel}) can be formally solved in terms of $\tilde{\phi}$, the formal inverse formula is practically useless because the independent kick model under dry friction is only valid for small $\Omega R_I/v_0$ and, thus the inverse Fourier transformation of Eq.~(\ref{sol_ik_bessel}) does not work well. 
Indeed, the exponential tail is reported for the dry frictional rotor for large $\Omega R_I/v_0$ in Ref. \cite{kssh2}, which cannot be captured by the independent kick solution.  

\subsection{Numerical implementation for the formula Eqs.~(\ref{sol_ik_bessel}) and (\ref{g_dry})}\label{dry_detail_sec}

\begin{figure}[h]
\begin{center}
\includegraphics[scale = 0.4]{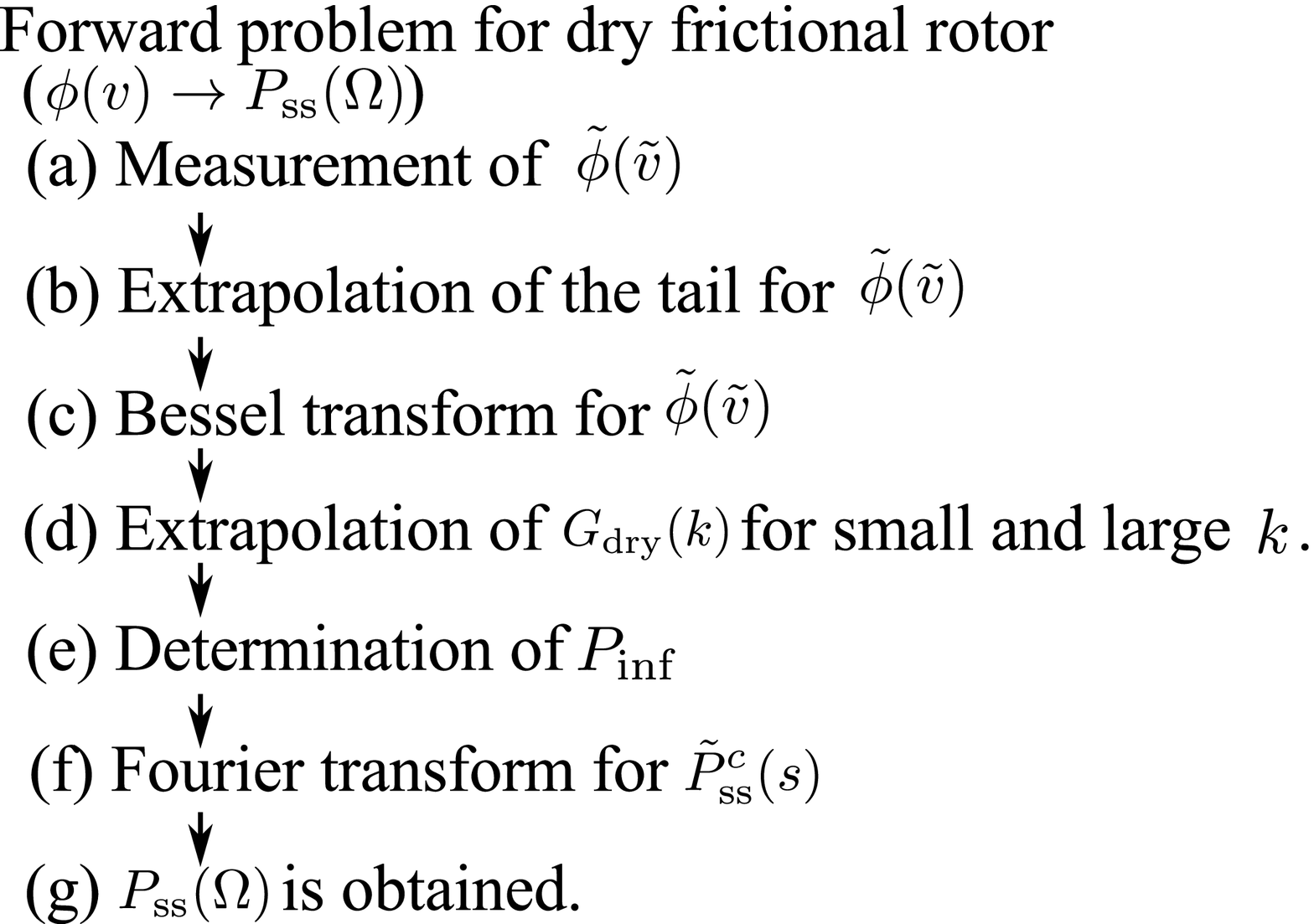}
\caption{(Color online) Outline of the forward problem for the dry rotor.}
\label{proc_forward_dry}
\end{center}
\end{figure}

In this section, a detailed technique to use the analytic PDF formulas (\ref{sol_ik_bessel}) and (\ref{g_dry}) is described. In a manner similar to Sec.~\ref{viscous_detail_sec}, this description is useful for both experimentalists and theoreticians interested in the details of the calculation. However, if readers are not interested in our technical details, they can skip this subsection. Fig.~\ref{proc_forward_dry} provides the outline of the procedure. The following six steps are involved (a)-(f) to obtain $P_{\rm ss}(\Omega)$ numerically.

(a) The granular velocity around the rotor is observed, and a histogram is obtained for the numerical VDF $\tilde{\phi}(\tilde{v})$. 

(b) A sufficiently large $\tilde{v}_-$ and $\tilde{v}_+ (\tilde{v}_- < \tilde{v}_+)$ are introduced to extrapolate the data for $\tilde{v}\tilde{\phi}(\tilde{v})$ in the $\tilde{v}\to\infty$ limit. Numerical $\tilde{v}\tilde{\phi}(\tilde{v})$ is fitted in the range $\tilde{v}_-<\tilde{v}<\tilde{v}_+$ with the exponential function $b_1 \exp(-b_2 \tilde{v})$, and the data are extrapolated for $\tilde{v}_-<\tilde{v}<\tilde{v}_{\rm end} = 20\tilde{v}_-$ by the fitting function. The cutoff for $\tilde{\phi}(\tilde{v})$ is introduced as $\tilde{v}_{\rm end} (> \tilde{v}_+)$. It is also verified that the following results are invariant if the Gaussian is used as the fitting function.

(c) Furthermore, $G_{\rm dry}(k)$ is obtained in terms of the Bessel transform Eq.~(\ref{g_dry}), and $C_1$ and $C_2$ are calculated according to Eq.~(\ref{c_n}).

(d) We interpolate the data of $G_{\rm dry}(k)$ for $0<k<k_{\rm cut} ^-$ by the fitting function $1/2\pi + C_1k^2 + C_2k^4$, which corresponds to the Taylor expansion of Eq.~(\ref{g_dry}) to avoid the numerical divergence of the second term in Eq.~(\ref{sol_ik_bessel}) in the $k\to 0$ limit. To extrapolate $G_{\rm dry}(k)$ for large $k$, we fit the data of $G_{\rm dry}(k)$ by $d_1\exp(-d_2 k^2)$ in the range $k_d ^-<k<k_d ^+$ with fitting parameters $d_1$ and $d_2$. We extrapolate the data using $d_1\exp(-d_2 k^2)$ in the region $k_d ^- < k<k_{\rm cut}^+$. $k_d ^-, k_d ^+,k_{\rm cut} ^-(<k_d ^-)$, and $k_{\rm cut} ^+ (> k_d ^+)$ are end points for the fitting ranges.

(e) We obtain $P_{\rm inf} = 1 + C_1\tilde{w}^2/\tilde{\Delta}'$ using Eq.~(\ref{c_n}).

(f) The Fourier transform is applied to the data for $\tilde{P}_{\rm ss} ^{c}(s)$ to obtain the smooth part of $P_{\rm ss}(\Omega)$ as $P_{\rm ss} ^{c}(\Omega) \equiv \int_{-\infty} ^{\infty} e^{-is\Omega}\tilde{P}_{\rm ss}^{c}(s)ds/2\pi$. 

(g) We obtain $P_{\rm ss}(\Omega) = P_{\rm inf}\delta(\Omega) + P_{\rm ss} ^{c}(\Omega)$. We note $P_{\rm ss}(\Omega = 0) = P_{\rm ss} ^{c}(0) + P_{\rm inf}/\Delta\Omega$. Here, $\Delta\Omega$ is the data mesh for $\Omega$. 
The fitting parameters introduced here are summarized in Table \ref{table_forward_dry}.

\begin{table}[h]
\caption{Summary of numerical values in the forward problem for the rotor under dry friction.}
\begin{tabular}{cc}
\hline\hline
symbols & values\\
\hline
$\tilde{v}_-$ & $15.0$\\
$\tilde{v}_+$ & $25.0$\\
$\tilde{v}_{\rm end}$ & $300.0$\\
$k_{\rm cut} ^-$ & $0.1$\\
$k_d ^-$ & $0.3$\\
$k_d ^+$ & $0.4$\\
$k_{\rm cut}^+$ & $0.35$\\
$b_1$ & $1182.192$\\
$b_2$ & $1.128878$\\
$d_1$ & $0.204485$\\
$d_2$ & $13.185175$\\
\hline\hline
\end{tabular}
\label{table_forward_dry}
\end{table}

\section{Summary and Discussion}\label{summary_sec}

We have examined the role of a granular rotor as a local non-equilibrium probe through the MD simulation of the rotor in vibrating granular beds under gravity. We have observed spatially inhomogeneous VDFs. We have formulated the inverse formula in cylindrical coordinates to explain the result of the MD simulation for a realistic viscous rotor. Starting from the Botlzmann-Lorentz equation, we have derived analytic formulas for the viscous frictional rotor. On the basis of the derived formulas, we have numerically calculated the angular VDF of the rotor from the data of VDF of the granular gas near the rotor, and vice versa. Furthermore, we have demonstrated that our inverse formula can be used even if the location of the rotor is different from the center of the container. Thus, the granular rotor is useful as a local probe for non-equilibrium baths.  

With respect to a rotor under dry friction, only the forward problem is considered, and the result corresponds to the MD result. The present study could not solve the inverse problem for the rotor under dry friction. In order to derive a valid inverse formula for the rotor under dry friction, it is expected that an appropriate interpolation between the independent kick solution and the exponential tail of the VDF of the rotor is necessary.

There are several possible extensions of this study. In the study, it is assumed that the restitution coefficient of grains is constant \cite{list_COR,list_COR1}. However, the restitution coefficient for a sphere depends on the impact velocity $v_{\rm imp}$ as $e(v_{\rm imp}) = 1 - {\mathcal B}_1v_{\rm imp} ^{1/5} + \cdots ({\mathcal B}_1> 0)$ \cite{brilliantov,kuwabara_1987,ramirez_1999}. As discussed in Ref. \cite{anna_brownian}, the velocity dependence of the restitution coefficient can be introduced to the Boltzmann-Lorentz equation, and this can be also analyzed by the theory proposed by the present study. It is necessary to analyze the effects of the tangential interaction and the mutual rotation between grains. It would be possible and interesting to apply our framework to the rotor in dense granular media \cite{camille} or denser granular liquids near the jamming transition beyond the Enskog equation \cite{suzuki_2015} by modifying the transition probability $W_{\epsilon}$~\cite{brey_self}. It is also particularly necessary for future studies to estimate the errors for the inverse estimation formulas.

\section*{Acknowledgement}
We are grateful for useful discussion with A. Puglisi and A. Gnoli. The numerical calculations were carried out on SR16000 at  YITP in Kyoto University. This work is supported by the Grants-in-Aid for Japan Society for Promotion of Science (JSPS) Fellows (Grants No.26$\cdot$2906, No.27$\cdot$6208, and No.28$\cdot$5315), and JSPS KAKENHI (Grant No.25287098 and 16H04025). This work is also partially supported by the JSPS core-to-core program for Nonequilibrium dynamics for soft matter and information.

\appendix

\section{Event driven simulation for rotor}\label{detail_rotor_MD}
This appendix explains the method to calculate the rotor dynamics in the event driven simulation. During the simulation, the system evolves with the time step of the minimum collision time $\Delta t^{\rm next}$. The collision time $\Delta t_{ij}$ between grains $i$ and $j$ $(i, j = 1,2,\cdots,N)$ can be precisely calculated, and the minimum of $\Delta t_{ij}$ can be obtained as $\Delta t^{\rm grain} = \min\{\Delta t_{ij}; i\ne j\}$ \cite{poschel}. The collision time $\Delta t_i ^{\rm wall}$ between the grain $i$ and the wall and the minimum can be exactly obtained as $\Delta t^{\rm wall} = \min\{\Delta t_i ^{\rm wall}; i\}$. However, collision times between the rotor and grains can not be calculated accurately. In the simulation, the collision time of grain $i$ with the rotor $\Delta t_i ^{{\rm rotor}}$ is approximately obtained by the virtual time evolution in a small time increment $1.0\times10^{-4}\sqrt{z_{\rm max}/2g}$. The collision time between a grain and the rotor is approximately calculated as the time when this distance becomes smaller than the radius of the grain after the virtual updating of the positions. If a grain collides with the wall before colliding with the rotor, it is considered that the grain does not collide with the rotor in this time step. Then, the minimum collision time between grains and the rotor can be obtained as $\Delta t^{\rm rotor} = {\rm min} \{\Delta t_i ^{\rm rotor}; i \}$. The time step of the simulation is obtained as $\Delta t^{\rm next} = \min\{\Delta t^{\rm grain}, \Delta t^{\rm wall}, \Delta t^{\rm rotor}\}$ by comparing three candidates of the collision time.

\section{Benchmark test for simulation}\label{benchmark}
\begin{figure}[h]
\begin{center}
\includegraphics[scale = 1.2]{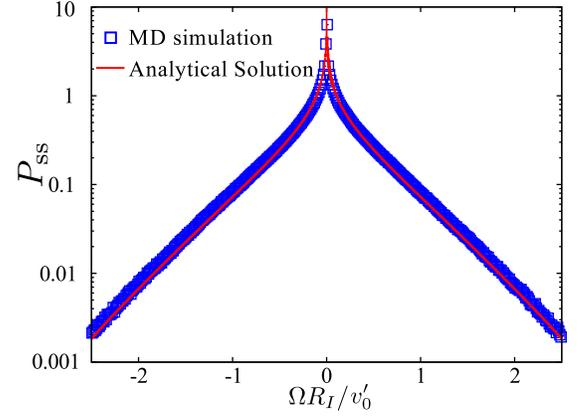}
\caption{(Color online) The results of the MD simulation under viscous friction (squares) and that obtained by the analytical solution proposed in this study are compared. The solid line represents the Fourier transform of Eq.~(\ref{pss_el}). The accurate solution derived in this study perfectly corresponds with the MD simulation data.} 
\label{pdf_el}
\end{center}
\end{figure}

In this appendix, it is shown that $P_{\rm ss}$ observed in the MD simulation can be analytically predicted when the gas is elastic $(e_{\rm g} = e_{\rm w} = e =1)$ without gravity $g=0$ and vibration. Therefore, this simple setup is examined under the condition $z_{\rm max} = 0$ as a benchmark test to examine the validity of the proposed method. A container heated by the thermal wall at the bottom is analyzed. The top and the side walls of the container are chosen to be smooth elastic walls. The post collisional velocity of a grain ${\bm v = (v_x,v_y,v_z)}$ against the thermal wall is selected as random, following the distribution 
$\phi_{\rm wall}({\bm v}, T_{\rm wall}) = \left({m}/{ T_{\rm wall}}\right)^2v_z \exp\left[-{m{\bm v} ^2}/{2T_{\rm wall}}\right]/{2\pi}$ \cite{maxwell1,maxwell2}.
The typical velocity in this setup $v_{0} ' = \sqrt{T_{\rm wall}/m}$ is selected instead of $v_0$ as shown in Sec.~\ref{setup_sec}. It is numerically verified that the VDF of the elastic gas can be considered as the Gaussian $\tilde{\phi}(\tilde{v}) = \phi_{\rm G}(\tilde{v})$ with $\tilde{v} = v/v_{0} '$. Furthermore, $P_{\rm ss}$ can be analytically obtained from the proposed theory. The obtained solution is compared with the MD simulation data in Fig.~\ref{pdf_el} for the viscous rotor.

By using Eqs.~(\ref{pss_analytical}) and (\ref{Psi_rotor}), the analytic solution for $P_{\rm ss}(\Omega)$ is obtained as
\begin{equation}
\tilde{P}_{\rm ss}\left(\frac{k}{\tilde{w}}\right) = \exp\left[ -\frac{1}{3\tilde{\gamma}\sqrt{2\pi}}\frac{k^2}{2} {}_2F_2\left( \begin{matrix} 1,1\\2,\frac{5}{2} \end{matrix}\left|-\frac{k^2}{2}\right. \right)\right]\label{pss_el},
\end{equation}
where $\tilde{\gamma}{B}={1}/{\sqrt{2\pi}}$, $\pi k \int_0 ^{\infty} dx x\phi_{\rm G}(x){\rm H}_0(kx)={k}D_{F}\left({k}/{\sqrt{2}}\right)/{\sqrt{\pi}}$, and
\begin{equation}
\int_0 ^{\frac{k}{\sqrt{2}}}\frac{ds'}{s'}\left(\frac{D_F(s')}{s'}-1 \right) = -\frac{k^2}{6}{}_2F_2\left( \begin{matrix} 1,1\\2,\frac{5}{2} \end{matrix}\left|-\frac{k^2}{2}\right. \right).
\end{equation}
Here, $D_F(x)$ denotes the Dawson function \cite{abramowitz}:
$D_F(x) \equiv e^{-x^2} \int_0 ^{x} e^{t^2} dt$
and ${}_qF_p$ denotes the generalized hypergeometric function \cite{abramowitz,andrews}:
\begin{eqnarray}
{}_qF_p\left(\left. \begin{matrix} a_1, a_2,\cdots a_q\\b_1,b_2,\cdots b_p \end{matrix}\right|z \right) \equiv \sum_{l = 0} ^{\infty} \frac{(a_1)_l (a_2)_l \cdots (a_q)_l}{(b_1)_l (b_2)_l \cdots (b_p)_l}\frac{z^l}{l!},
\end{eqnarray}
where the Pochhammer symbol is introduced as $(a)_l \equiv \Gamma(a+l)/\Gamma(a)$ with $l\geq 0$. Here, $\Gamma(a)$ represents the Gamma function $\Gamma(a) = \int_0 ^{\infty} s^{a-1}e^{-s}ds$. In order to plot Fig.~\ref{pdf_el}, $\gamma /Mz_{\rm max}v_0 = 0.50$, which corresponds to $\tilde{\gamma} = 0.57624$ is adopted. The histogram for the angular velocity is shown in Fig.~\ref{pdf_el} and the analytical solution (solid line) is consistent with the MD simulation result (squares) thereby ensuring the accuracy of the MD code and the validity of the proposed framework.

\section{Detailed derivation of the formulas for a rotor under viscous friction}\label{app_vis}
In this appendix, the detailed derivation of the analytic results for the viscous frictional rotor in Sec.~\ref{vis_sec} is shown. 
Specifically, given that ${\mathcal Y} \equiv y/\epsilon$ is introduced, the transition rate given by ${\mathcal W}({\mathcal Y}) \equiv \bar{W}(\omega = 0; {\mathcal Y})$ is independent of $\epsilon$ and $\Omega$. It is noted that $W_{\epsilon}(\omega;y)$ satisfies the relation $\bar{W}(\omega;{\mathcal Y})d{\mathcal Y} = W_{\epsilon}(\omega;y)dy$ up to the leading order. We obtain the reduced time evolution equation for ${\mathcal P} = {\mathcal P}(\Omega,t)$ in $\epsilon\to 0$ limit from Eq.~(\ref{bl_eq}) when the axial friction is sufficiently strong according to the generalized system size expansion \cite{kssh1,kssh2}. For general frictional torque $N_{\rm fri} = N_{\rm fri}(\omega)$, the following expression is obtained:
\begin{eqnarray}
\frac{\partial {\mathcal P}}{\partial t} &=& -\frac{1}{I}\left\{\frac{\partial}{\partial\Omega}\tilde{N}_{\rm fri}(\Omega){\mathcal P}\right\} \nonumber\\ &&+ \int_{-\infty} ^{\infty} d{\mathcal Y}{\mathcal W}({\mathcal Y})\left\{{\mathcal P}(\Omega - {\mathcal Y},t)- {\mathcal P}(\Omega, t)\right\},\nonumber\\ \label{bl_scaled}\\
{\mathcal W}(\mathcal Y) &=& \rho h \int_0 ^{2w} d{\sigma}\int_{-\infty} ^{\infty}dv_xdv_y\phi(v_x,v_y)\Theta(-{v}_n)\nonumber\\
&& \times |-{v}_n|\delta\left({\mathcal Y} -\Delta\bar{\omega}'\right),\\
\Delta\bar{\omega}' &\equiv& -\frac{1+e}{R_I}g(\sigma)v_n\label{delta_omega},
\end{eqnarray}
where the rescaled friction $\tilde{N}_{\rm fri}(\Omega) = N_{\rm fri}(\epsilon\Omega)/\epsilon$ is introduced. Additionally, $\tilde{N}_{\rm fri}(\Omega) = O(1)$ is assumed in $\epsilon\to 0$ limit.

The scaled friction is expressed as $\tilde{N}_{\rm fri} = -\tilde{\gamma}\Omega$ in Eq.~(\ref{bl_scaled}) for the case of viscous friction with $\tilde{\gamma} = \gamma/(2\rho hwIv_0)$.
The cumulant generating function $\Phi(s)$ is calculated. For an even integer $l$, the cumulant ${\mathcal K}_l = \int d{\mathcal Y}{\mathcal Y}^l{\mathcal W}(\mathcal Y)$ is calculated as follows:
\begin{eqnarray}
{\mathcal K}_l &=& \rho h \int_0 ^{2w} d{\sigma}\int_{-\infty} ^{\infty}dv_xdv_y\phi(v_x,v_y)\nonumber\\
&&\times\Theta(-{v}_n) |-{v}_n| (\Delta\bar{\omega}')^l\nonumber\\
&=& \rho h\frac{(1+e)^l}{R_I ^l} 2\int_{-w/2} ^{w/2} d{\sigma'} \left(\frac{{\sigma'}}{R_I}\right)^l\nonumber\\
&& \times\int_0 ^{\infty}dv v \int_0 ^{2\pi}d\theta\phi(v)\Theta(-v\cos\theta)(-v\cos\theta)^{l+1}\nonumber\\
&=&  \rho h\frac{(1+e)^l}{R_I ^l}4\int_{0} ^{w/2}d{\sigma'} \left(\frac{{\sigma'}}{R_I}\right)^l\nonumber\\
&& \times\int_0 ^{\infty}dv v^{l+2}\phi(v)\frac{\Gamma\left(\frac{l+2}{2}\right)}{\Gamma\left(\frac{l+3}{2}\right)}\sqrt{\pi}\nonumber\\
&=&\rho h\frac{\sqrt{\pi}(1+e)^l}{R_I ^{2l}}\frac{\Gamma\left(\frac{l+2}{2}\right)}{\Gamma\left(\frac{l+3}{2}\right)}\nonumber\\
&&\times\frac{4}{l+1}\left(\frac{w}{2}\right)^{l+1}\int_0 ^{\infty} dvv^{l+2}\phi(v).
\end{eqnarray}
The coordinate variable $\sigma$ to $\sigma'$ satisfying $-w/2<\sigma'<w/2$, is changed, and the front-back symmetry is used for the rotor. Then, $\Phi(s)$ is written as follows:
\begin{eqnarray}
\Phi(s) &=& \int_0 ^{\infty} dv\phi(v)2\sqrt{\pi}\rho h\sum_{l=1} ^{\infty} \frac{w^{l+1}v^{l+2}}{R_I ^{2l} 2^l}\nonumber\\ 
&&\times\frac{(is)^l(1+e)^l}{(l+1)!}\frac{\Gamma\left(\frac{l+2}{2}\right)}{\Gamma\left(\frac{l+3}{2}\right)}\nonumber\\
&=& - \int_0 ^{\infty} dv\phi(v)2\sqrt{\pi}\rho hw \frac{R_I ^4}{w^2 s^2}\left(\frac{2}{1+e} \right)^2 \nonumber\\ 
&&\times\left\{\sum_{j = 2} ^{\infty}\frac{(-1)^j}{(2j-1)!}\frac{\Gamma(j)}{\Gamma\left(j+\frac{1}{2}\right)} \left(\frac{1+e}{2}\frac{wsv}{R_I ^2} \right)^{2j}\right\}\nonumber\\
&=&  - \int_0 ^{\infty} d\tilde{v}\tilde{\phi}(\tilde{v}) \frac{2\rho hwv_0}{\tilde{s}^2\tilde{w}^2} \nonumber\\
&&\times\left\{-(\tilde{w}\tilde{s}\tilde{v})\pi{\rm H}_0(\tilde{w}\tilde{s}\tilde{v}) +2(\tilde{w}\tilde{s}\tilde{v})^2\right\},
\end{eqnarray}
where the Struve function is introduced as follows: 
\begin{eqnarray}
{\rm H}_0(y) &=& \sum_{l=0} ^{\infty}\frac{(-1)^ly^{2l+1}}{\{(2l+1)!!\}^2}\label{struve0_def}.
\end{eqnarray}
Because $\Phi(s)$ and $\tilde{P}_{\rm ss}$ satisfy the relation
\begin{eqnarray}
\Phi(s) = s\frac{\gamma}{I} \frac{d}{ds}{\rm ln}\tilde{P}_{\rm ss},
\end{eqnarray}
we obtain
\begin{eqnarray}
\tilde{s}^3\frac{d}{d\tilde{s}}{\rm ln}\tilde{P}_{\rm ss} &=& \frac{2\rho hwIv_0}{\gamma\tilde{w}^2} \int_0 ^{\infty} d\tilde{v}\tilde{\phi}(\tilde{v})\nonumber\\
&&\times\left\{(\tilde{w}\tilde{s}\tilde{v})\pi{\rm H}_0(\tilde{w}\tilde{s}\tilde{v}) -2(\tilde{w}\tilde{s}\tilde{v})^2\right\}.
\end{eqnarray}
The variable $k = \tilde{w}\tilde{s}$ is introduced to obtain the following expression:
\begin{eqnarray}
\tilde{\gamma}\left\{k^3 \frac{d}{dk}{\rm ln}\tilde{P}_{\rm ss}\left(\frac{k}{\tilde{w}}\right)+Bk^2\right\} = \pi k\int_0 ^{\infty}d\tilde{v} \tilde{v}\tilde{\phi}(\tilde{v}){\rm H}_0(k\tilde{v}),\nonumber\\ \label{eq_last_app}
\end{eqnarray}
where $B = (2/\tilde{\gamma})\int_0 ^{\infty}d\tilde{v} \tilde{v}^2 \tilde{\phi}(\tilde{v})$ is introduced. Thus, the formula (\ref{eq_last_main}) is obtained.

\section{Detailed derivation of the formulas for a rotor under dry friction}\label{app_dry}
This appendix shows the detailed derivation of the analytic results for a rotor under dry friction in Sec.~\ref{dry_sec}. The derivation is initialized by using Eqs.~(\ref{bl_scaled})-(\ref{delta_omega}) with the scaled dry friction $\tilde{N}_{\rm fri}(\Omega) = -\tilde{\Delta}{\rm sgn}(\Omega)$. The Fourier transform is applied to the equation $\partial {\mathcal P}/\partial t =0$ to obtain the independent kick solution at the order $O(1/\tilde{\Delta}')$ as in Eq.~(\ref{sol_ik}). Equation (\ref{sol_ik}) is rewritten as follows:
\begin{eqnarray}
\frac{\tilde{\Delta}}{I}(\tilde{P}_{\rm ss} - 1) &=& \int_{-\infty} ^{\infty}d{\mathcal Y} {\mathcal W}({\mathcal Y})\int_0 ^{\mathcal Y} \frac{d\Omega}{{\rm sgn}(\Omega)}\sum_{l = 1} ^{\infty} \frac{(is\Omega)^l}{l!}\nonumber\\
&=&  \sum_{l = 1} ^{\infty} \frac{(is)^l}{l!}\int_{-\infty} ^{\infty}d{\mathcal Y} {\mathcal W}({\mathcal Y})\frac{{\rm sgn}({\mathcal Y}){\mathcal Y}^{l+1} }{l+1}\nonumber\\
&=&\sum_{j = 1} ^{\infty} \frac{(-1)^j s^{2j+2}}{(2j+1)!}\int_{-\infty} ^{\infty} d{\mathcal Y}|{\mathcal Y}|{\mathcal Y}^{2j}{\mathcal W}(\mathcal Y)\frac{1}{s^2}.\nonumber\\
\end{eqnarray}
Given Eqs. (\ref{l_relation1})-(\ref{l_relation4}) for an even integer $l$:
\begin{widetext}
\begin{eqnarray}
\int_{-\infty} ^{\infty} d{\mathcal Y}|{\mathcal Y}|^{l+1}{\mathcal W}(\mathcal Y) &=&\frac{4\rho h(1+e)^{l+1}}{R_I ^{2l+2}(l+2)}\left(\frac{w}{2}\right)^{l+2}\int_0 ^{\infty} dvv^{l+3}\phi(v)\int_{-\pi/2}^{\pi/2}d\theta \cos^{l+2}\theta\nonumber\\
&=&\frac{4\rho h(1+e)^{l+1}\sqrt{\pi}}{R_I ^{2l+2}(l+2)}\left(\frac{w}{2}\right)^{l+2}\frac{l+4}{l+3}\frac{\left(\frac{l}{2}+\frac{3}{2}\right)!}{\left(\frac{l}{2}+2\right)!}\int_0 ^{\infty} dvv^{l+3}\phi(v)\label{l_relation1},
\end{eqnarray}
\end{widetext}
\begin{eqnarray}
\int_{-\pi/2}^{\pi/2}d\theta \cos^{l+2}\theta&=&\sqrt{\pi}\frac{\Gamma\left(\frac{l+3}{2}\right)}{\Gamma\left(\frac{l+4}{2}\right)}\nonumber\\
&=& \frac{l+4}{l+3}\frac{\left(\frac{l}{2}+\frac{3}{2}\right)!}{\left(\frac{l}{2}+2\right)!},\label{l_relation2}\\
\frac{(2j+4)\left(j+\frac{3}{2}\right)!}{(2j+3)!(j+2)!}&=&\sqrt{\pi}\left(\frac{1}{2^{j+1}(j+1)!}\right)^2,\label{l_relation3}\\
J_0(x) &=& 1 - \sum_{j=0} ^{\infty}\frac{(-1)^jx^{2(j+1)}}{2^{2(j+1)}\{(j+1)!\}^2},\nonumber\\ \label{l_relation4}
\end{eqnarray}
we obtain the following expression:
\begin{eqnarray}
\frac{\tilde{\Delta}}{I}\left\{\tilde{P}_{\rm ss}\left(\frac{k}{\tilde{w}}\right) - 1\right\}\left(\frac{k}{\tilde{w}}\right)^2 &=&\frac{4{\pi} \rho hv_0 ^2}{1+e}\int_0 ^{\infty}d\tilde{v} \tilde{v}\tilde{\phi}(\tilde{v})\nonumber\\ 
&&\times\left\{1 - \frac{k^2\tilde{v}^2}{4} - J_0(k\tilde{v})\right\}.\nonumber\\ \label{result_dry_app}
\end{eqnarray}
Hence, Eqs.~(\ref{sol_ik_bessel}) and (\ref{g_dry}) are obtained from Eq.~(\ref{result_dry_app}).


\begin{thebibliography}{2}
\bibitem{eng1} B. J. Ennis, J. Green, and R. Davis, Particle Technol. {\bf 90}, 32. (1994).
\bibitem{eng2} T. M. Knowlton, J. W. Carson, G. E. Klinzing, and W.-C. Yang, 1994, Particle Technol. {\bf 90}, 44. (1994).
\bibitem{nagel} H. M. Jaeger, S. R. Nagel, and R. P. Behringer, {Granular solids, liquids, and gases}, Rev. Mod. Phys. {\bf 68}, 1259 (1996).
\bibitem{aranson} I. S. Aranson and L. S. Tsimring, {Patterns and collective behavior in granular media: Theoretical concepts}, Rev. Mod. Phys. {\bf 78}, 641 (2006).
\bibitem{poschel} T. P\"oschel and T. Schwager, {\itshape Computational Granular Dynamics} (Springer, Berlin, 2005).
\bibitem{brilliantov} N. V. Brilliantov and T. P\"oschel, {\itshape Kinetic Theory of Granular Gases} (Oxford University Press, New York, 2010).

\bibitem{goldshtein_1995} A. Goldshtein and M. Shapiro, {Mechanics of collisional motion of granular materials. Part 1. General hydrodynamic equations}, J. Fluid Mech. {\bf 282}, 75 (1995).
\bibitem{esipov_1997} S. E. Esipov and T. P\"oschel, {The granular phase diagram}, J. Stat. Phys. {\bf 86}, 1385 (1997).
\bibitem{van_noije_1998} T.P.C. van Noije and M.H. Ernst, {Velocity distributions in homogeneous granular fluids: the free and the heated case}, Granul. Matt. {\bf 1}, {57} (1998). 
\bibitem{olafsen1} J. S. Olafsen and J. S. Urbach, {Velocity distributions and density fluctuations in a granular gas}, Phys. Rev. E {\bf 60}, R2468(R) (1999).
\bibitem{brilliantov_2000} N. V. Brilliantov, T. P\"oschel, {Velocity distribution in granular gases of viscoelastic particles}, Phys. Rev. E {\bf 61}, 5573 (2000).
\bibitem{kudroli} A. Kudrolli and J. Henry, {Non-Gaussian velocity distributions in excited granular matter in the absence of clustering}, Phys. Rev. E {\bf 62}, R1489(R) (2000).
\bibitem{olafsen2} G. W. Baxter and J. S. Olafsen, {Kinetics: Gaussian statistics in granular gases}, Nature {\bf 425}, 680 (2003).
\bibitem{kawarada} A. Kawarada and H. Hayakawa, {Non-Gaussian Velocity Distribution Function in a Vibrating Granular Bed}, J. Phys. Soc. Jpn. {\bf 73}, 2037-2040 (2004).
\bibitem{dubey_2013} A. K. Dubey, A. Bodrova, S. Puri, N. Brilliantov, {Velocity distribution function and effective restitution coefficient for a granular gas of viscoelastic particles}, Phys. Rev. E {\bf 87}, 062202 (2013).

\bibitem{jenkins_savage_1983} J. T. Jenkins and S. B. Savege, {A theory for the rapid flow of identical, smooth, nearly elastic, spherical particles}, J. Fluid. Mech. {\bf 130}, 187 (1983).
\bibitem{jenkins_richman_1985} J. T. Jenkins and M. W. Richman, {Kinetic theory for plane shear flows of a dense gas of identical, rough, inelastic, circular disks}, Phys. Fluids. {\bf 28}, 3485 (1985).
\bibitem{garzo_dufty} V. Garz\'o and J. W. Dufty, {Dense fluid transport for inelastic hard spheres}, Phys. Rev. E {\bf 59}, 5895 (1999).
\bibitem{lutsko} J. F. Lutsko, {Transport properties of dense dissipative hard-sphere fluids for arbitrary energy loss models}, Phys. Rev. E {\bf 72}, 021306 (2005).
\bibitem{hcs} C. C. Maa\ss, N. Isert, G. Maret, and C. M. Aegerter, {Experimental Investigation of the Freely Cooling Granular Gas}, Phys. Rev. Lett. {\bf 100}, 248001 (2008).
\bibitem{micro_g} {\rm e.g.} S. Tatsumi, Y. Murayama, H. Hayakawa, and M. Sano, {Experimental study on the kinetics of granular gases under microgravity}, J. Fluid. Mech. {\bf 641}, 521 (2009).
\bibitem{gdrmidi} GDR MiDi, {On dense granular flows}, Eur. Phys. J. E {\bf 14}, 341 (2004).
\bibitem{forterre} Y. Forterre and O. Pouliquen, {Flows of Dense Granular Media}, Annu. Rev. Fluid Mech. {\bf 40}, 1 (2008).
\bibitem{jet1} T. G. Sano and H. Hayakawa, {Simulation of granular jets: Is granular flow really a perfect fluid?}, Phys. Rev. E {\bf 86}, 041308 (2012).
\bibitem{jet2} T. G. Sano and H. Hayakawa, {Numerical analysis of impact processes of granular jets}, AIP Conf. Proc. {\bf 1542}, 622 (2013).
\bibitem{jet3} T. G. Sano and H. Hayakawa, {Jet-induced jammed states of granular jet impacts}, Prog. Theor. Exp. Phys. ({\bf 2013}) 103J02.
\bibitem{cheng2} X. Cheng, L. Gordillo, W. W. Zhang, H. M. Jaeger, and S. R. Nagel, {Impact dynamics of granular jets with noncircular cross sections}, Phys. Rev. E {\bf 89}, 042201 (2014).

\bibitem{jam1} A. J. Liu and S. R. Nagel, {Nonlinear dynamics: Jamming is not just cool any more}, Nature {\bf 396}, 21 (1998).
\bibitem{jam2} M. Otsuki and H. Hayakawa, {Critical scaling near jamming transition for frictional granular particles}, Phys. Rev. E {\bf 83}, 051301 (2011).
\bibitem{hayakawa_2013} H. Hayakawa and M. Otsuki, {Nonequilibrium identities and response theory for dissipative particles}, Phys. Rev. E {\bf 88}, 032117 (2013).
\bibitem{suzuki_2015} K. Suzuki and H. Hayakawa, {Divergence of Viscosity in Jammed Granular Materials: A Theoretical Approach}, Phys. Rev. Lett. {\bf 115}, 098001 (2015).

\bibitem{mri} E. E. Ehrichs, et al., {Granular Convection Observed by Magnetic Resonance Imaging}, Science {\bf 267} 1632 (1995).
\bibitem{imaging} A. V. Orpe and A. Kudroli, {Velocity correlations in dense granular flows observed with internal imaging}, Phys. Rev. Lett. {\bf 98} 238001 (2007).

\bibitem{brey} J. J. Brey, M. J. Ruiz-Montero, and R. Garc\'ia-Rojo, {Brownian motion in a granular gas}, Phys. Rev. E {\bf 60}, 7174 (1999).
\bibitem{saraccino} A. Sarracino, D. Villamaina, G. Costantini, and A. Puglisi, {Granular Brownian motion}, J. Stat. Mech. ({\bf 2010}) P04013.

\bibitem{naert1} A. Naert, {Experimental study of work exchange with a granular gas: The viewpoint of the Fluctuation Theorem}, Europhys. Lett. {\bf 97}, 20010 (2012).
\bibitem{devaraj} S. Joubaud, D. Lohse, and D. van der Meer, {Fluctuation Theorems for an Asymmetric Rotor in a Granular Gas}, Phys. Rev. Lett. {\bf 108}, 210604 (2012).
\bibitem{gnoli1} A. Gnoli, A. Puglisi, and H. Touchette, {Granular Brownian motion with dry friction}, Europhys. Lett. {\bf 102}, 14002 (2013).
\bibitem{gnoli2} A. Gnoli, A. Sarracino, A. Puglisi, and A. Petri, {Nonequilibrium fluctuations in a frictional granular motor: Experiments and kinetic theory}, Phys. Rev. E {\bf 87}, 052209 (2013).
\bibitem{gnoli3} A. Gnoli, A. Petri, F. Dalton, G. Pontuale, G. Gradenigo, A. Sarracino, and A. Puglisi, {Brownian Ratchet in a Thermal Bath Driven by Coulomb Friction}, Phys. Rev. Lett. {\bf 110}, 120601 (2013).
\bibitem{naert2}  C.-\'E. Lecomte and A. Naert, {Experimental study of energy transport between two granular gas thermostats}, J. Stat. Mech. ({\bf 2014}) P11004.
\bibitem{loreto_meer_2016} L. O. G\'alvez and D. van der Meer, {Granular motor in the non-Brownian limit}, J. Stat. Mech. ({\bf 2016}) 043206.


\bibitem{kssh1} K. Kanazawa, T. G. Sano, T. Sagawa, and H. Hayakawa, {Minimal Model of Stochastic Athermal Systems: Origin of Non-Gaussian Noise}, Phys. Rev. Lett. {\bf 114}, 090601 (2015).
\bibitem{kssh2} K. Kanazawa, T. G. Sano, T. Sagawa, and H. Hayakawa, {Asymptotic Derivation of Langevin-like Equation with Non-Gaussian Noise and Its Analytical Solution}, J. Stat. Phys. {\bf 160}, 1294 (2015).

\bibitem{cleuren} B. Cleuren and R. Eichhorn, {Dynamical properties of granular rotors}, J. Stat. Mech. ({\bf 2008}) P10011.
\bibitem{talbot} J. Talbot, R. D. Wildman, and P. Viot, {Kinetics of a Frictional Granular Motor}, Phys. Rev. Lett. {\bf 107}, 138001 (2011).



\bibitem{smolchowski} M. Smoluchowski, {Zur kinetischen Theorie der Brownschen Molekularbewegung und der Suspensionen}, Annalen der Physik {\bf 21}, 756 (1906).
\bibitem{chapman} S. Chapman, {On the Brownian displacements and thermal diffusion of grains suspended in a non-uniform fluid}, Proc. Roy. Soc. Ser. A {\bf 119}, 34 (1928).
\bibitem{kolmogorov} A. Kolmogorov, {Ueber die analytischen Methoden in der Wahrscheinlichkeitsrechnung}, Math. Ann. {\bf 104}, 415 (1931).
\bibitem{kampen} N. G. van Kampen, {\itshape Stochastic Processes in Physics and Chemistry}, 3rd ed. (North-Holland Personal Library, 2007).
\bibitem{gardiner} C. Gardiner, {\itshape Stochastic Methods, 4th ed.} (Springer-Verlag, Berlin, 2009).

\bibitem{list_COR} A. Lorenz, C. Tuozzolo, and M.Y. Louge, {Measurements of impact properties of small, nearly spherical particles}, Exp. Mech. {\bf 37}, 292 (1997). 
\bibitem{list_COR1} M. Louge (1999), {Fall 1999 Impact Parameter Chart}, \url{http://grainflowresearch.mae.cornell.edu/impact/data/Impact%20Results.html}
\bibitem{jenkins_zhang_2002} J. T. Jenkins and C. Zhang, {Kinetic theory for identical, frictional, nearly elastic spheres}, Phys. Fluids {\bf 14}, 1228 (2002). 
\bibitem{yoon_jenkins_2005} D. K. Yoon and J. T. Jenkins, {Kinetic theory for identical, frictional, nearly elastic disks}, Phys. Fluids {\bf 17}, 083301 (2005).
\bibitem{saitoh_hayakawa_2007} K. Saitoh and H. Hayakawa, {Rheology of a granular gas under a plane shear}, Phys. Rev. E {\bf 75}, 021302 (2007).
\bibitem{jenkins_2010} J. T. Jenkins and D. Berzi, {Dense inclined flows of inelastic spheres: tests of an extension of kinetic theory}, Granul. Matt. {\bf 12}, 151 (2010).
\bibitem{maxwell1} Y. Sone, {\itshape Molecular Gas Dynamics: theory, techniques, and applications} (Birkh\"auser, Boston, 2007)
\bibitem{maxwell2} T. G. Sano and H. Hayakawa, {Efficiency at maximum power output for an engine with a passive piston}, Progr. Theor. Exp. Phys., {\itshape in press} (arXiv: 1412.4468).

\bibitem{sandpaper} S-S. Hsiau and W-L. Yang, {Stresses and transport phenomena in sheared granular flows with different wall
conditions}, Phys. Fluids. {\bf 14}, 612 (2002).

\bibitem{pise_wise_ref} D. van der Meer and P. Reimann, {Temperature anisotropy in a driven granular gas}, Europhys. Lett. {\bf 74}, 384 (2006).


\bibitem{evans_morris} D. J. Evans and G. P. Morriss, {\itshape Statistical Mechanics of Nonequilibrium Liquids} (Academic Press, London, 1990).
\bibitem{puglisi_1998} A. Puglisi, V. Loreto, U. Marini Bettolo Marconi, A. Petri, and A. Vulpiani, {Clustering and Non-Gaussian Behavior in Granular Matter} Phys. Rev. Lett. {\bf 81}, 3848 (1998).
\bibitem{van_noije_1999} T. P. C. van Noije, M. H. Ernst, E. Trizac, and I. Pagonabarraga, {Randomly driven granular fluids: Large-scale structure}, Phys. Rev. E {\bf 59}, 4326 (1999).
\bibitem{gradenigo_2011} G. Gradenigo, A. Sarracino, D. Villamaina, and A. Puglisi, {Non-equilibrium length in granular fluids: From experiment to fluctuating hydrodynamics}, Europhys. Lett. {\bf 96}, 14004 (2011).
\bibitem{khalil_2014} N. Khalil and V. Garz\'o, {Homogeneous states in driven granular mixtures: Enskog kinetic theory versus molecular dynamics simulations}, J. Chem. Phys. {\bf 140}, 164901 (2014).
\bibitem{persson} B. N. J. Persson, {\itshape Sliding Friction} (Springer-Verlag, Berlin, 2000).
\bibitem{friction_h} H. Hayakawa, {Langevin equation with Coulomb friction}, Physica D {\bf 205}, 48 (2005).
\bibitem{genne} P. G. de Gennes, {Brownian Motion with Dry Friction}, J. Stat. Phys. {\bf 119}, 953 (2005).
\bibitem{friction_review} T. Baumberger and C. Caroli, {Solid friction from stick-slip down to pinning and aging}, Adv. Phys {\bf 55}, 279 (2006). 
\bibitem{sano_hayakawa} T. G. Sano and H. Hayakawa, {Roles of dry friction in the fluctuating motion of an adiabatic piston}, Phys. Rev. E {\bf 89}, 032104 (2014).

\bibitem{eliazar} I. Eliazar and J. Klafter, {L\'evy-Driven Langevin Systems: Targeted Stochasticity}, J. Stat. Phys. {\bf 111}, 739 (2003).
\bibitem{abramowitz} M. Abramowitz and I. A. Stegun, {\itshape Handbook of Mathematical Functions: With Formulas, Graphs, and Mathematical Tables} (Dover, New York, 1964).

\bibitem{titchmarsh} E. C. Titchmarsh, {A pair of inversion formulas}, Proc. London Math. Soc. {\bf 2}, 23 (1923).

\bibitem{kuwabara_1987} G. Kuwabara and K. Kono, {Restitution Coefficient in a Collision between Two Spheres}, Jpn. J. Appl. Phys. {\bf 26}, 1230 (1987).
\bibitem{ramirez_1999} R. Ramirez, T. P\"oschel, N. V. Brilliantov, T. Schwager, {Coefficient of restitution of colliding viscoelastic spheres}, Phys. Rev. E {\bf 60} 4465 (1999).

\bibitem{double_exponential} H. Takahasi and M. Mori, Double Exponential Formulas for Numerical Integration, {\itshape Publications of RIMS, Kyoto University} {\bf 9}, 721 (1974).
\bibitem{andrews} G. E. Andrews, R. Askey, and R. Roy, {\itshape Special functions} (Cambridge University Press, Cambridge, England, 1999).

\bibitem{anna_brownian} A. S. Bodrova, N. V. Brilliantov, and A. Yu. Loskutov, {Brownian motion in granular gases of viscoelastic particles }, J. Exp. Theor. Phys. {\bf 109}, 946 (2009).

\bibitem{camille} C. Scalliet, A. Gnoli, A. Puglisi, and A. Vulpiani, {Cages and Anomalous Diffusion in Vibrated Dense Granular Media}, Phys. Rev. Lett. {\bf 114}, 198001 (2015).
\bibitem{brey_self} J. Javier Brey, M. J. Ruiz-Montero, D. Cubero, and R. Garc\'ia-Rojo, {Self-diffusion in freely evolving granular gases}, Phys. Fluids {\bf 12}, 876 (2000).
\end{thebibliography}
\end{document}